\documentclass[prx,twocolumn,superscriptaddress,longbibliography]{revtex4-2}

\newif\ifNotDraft                   
\NotDrafttrue                      

\usepackage{graphicx}               
\usepackage{pgfplots}               
\pgfplotsset{compat=newest}         
\usepackage{placeins}               
\usepackage{amsmath}                
\usepackage{xfrac}                  

\definecolor{customlinkcolor}{RGB}{46,48,146}

\definecolor{my_yellow}{RGB}{255,180,0}
\definecolor{my_green}{RGB}{130,200,0}

\definecolor{my_light_blue}{RGB}{128,128,255}
\definecolor{my_lighter_blue}{RGB}{215,215,255}

\definecolor{my_light_red}{RGB}{255,128,128}
\definecolor{my_lighter_red}{RGB}{255,215,215}

\usepackage[colorlinks=true,allcolors=customlinkcolor]{hyperref}    

\DeclareRobustCommand{\orcidicon}{%
    \hspace{-1.1mm}                             
    \includegraphics[width=2.5mm]{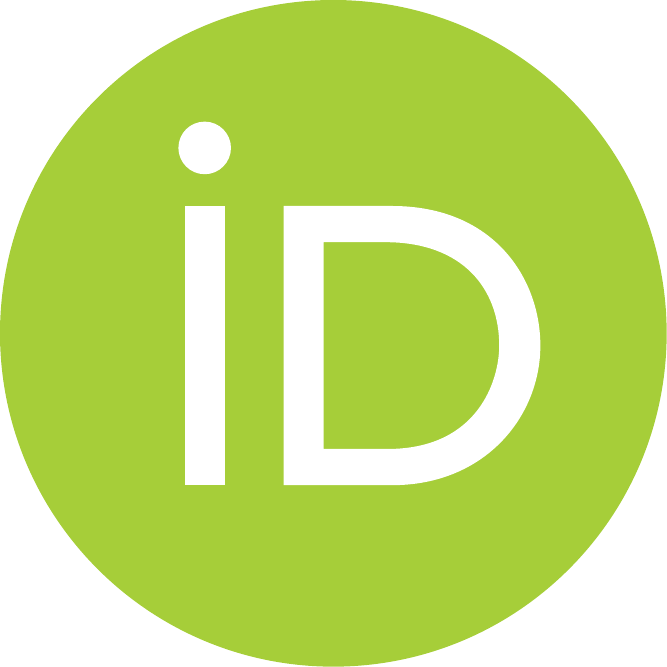} 
    \hspace{-1.3mm}}                            
\newcommand{\orcid}[1]{\href{https://orcid.org/#1}{\orcidicon}}
    \usepackage[caption=false,justification=justified]{subfig}
    \usepackage{ragged2e}
    \DeclareCaptionJustification{justified}{\justifying}
    
    \usepackage{mathtools}
    
    \DeclarePairedDelimiter\ket{\lvert}{\rangle}
    \DeclarePairedDelimiterX\braket[2]{\langle}{\rangle}{#1 \delimsize\vert #2} 

    \usepackage{cleveref}
    \crefname{equation}{Eq.}{Eqs.}
    \crefname{figure}{Fig.}{Figs.}
    \crefname{table}{Tab.}{Tabs.}

\begin{document}

\title{Precision Measurement of the Excited State Landé g-factor and Diamagnetic Shift of the Cesium D\textsubscript{2} Line}

\author{Hans Stærkind\orcid{0000-0002-6844-3305}}
\email{hans.staerkind@nbi.ku.dk}
\affiliation{Niels Bohr Institute, University of Copenhagen, Blegdamsvej 17, 2100 Copenhagen, Denmark}
\affiliation{Danish Research Centre for Magnetic Resonance, Centre for Functional and Diagnostic Imaging and Research, Copenhagen University Hospital - Amager and Hvidovre, Kettegard Allé 30, 2650 Hvidovre, Denmark}
\author{Kasper Jensen\orcid{0000-0002-8417-4328}}
\affiliation{Niels Bohr Institute, University of Copenhagen, Blegdamsvej 17, 2100 Copenhagen, Denmark}
\affiliation{School of Physics and Astronomy, University of Nottingham, University Park, Nottingham NG7 2RD, England, United Kingdom}
\author{Jörg H. Müller\orcid{0000-0001-6984-0487}}
\affiliation{Niels Bohr Institute, University of Copenhagen, Blegdamsvej 17, 2100 Copenhagen, Denmark}
\author{Vincent~O.~Boer\orcid{0000-0001-6026-3134}}
\affiliation{Danish Research Centre for Magnetic Resonance, Centre for Functional and Diagnostic Imaging and Research, Copenhagen University Hospital - Amager and Hvidovre, Kettegard Allé 30, 2650 Hvidovre, Denmark}
\author{Esben Thade Petersen\orcid{0000-0001-7529-3432}}
\affiliation{Danish Research Centre for Magnetic Resonance, Centre for Functional and Diagnostic Imaging and Research, Copenhagen University Hospital - Amager and Hvidovre, Kettegard Allé 30, 2650 Hvidovre, Denmark}
\affiliation{Section for Magnetic Resonance, DTU-Health Tech, Technical University of Denmark, Oersteds Plads, Building 349, 1st floor, 2800 Kgs Lyngby, Denmark}
\author{Eugene S. Polzik\orcid{0000-0001-9859-6591}}
\affiliation{Niels Bohr Institute, University of Copenhagen, Blegdamsvej 17, 2100 Copenhagen, Denmark}

\date{\today}

\begin{abstract}
We have performed saturated absorption spectroscopy on the cesium D\textsubscript{2} line in 3 T and 7 T magnetic fields. By means of sideband spectroscopy on the extreme angular momentum states we have measured the linear magnetic frequency shift of the transition to be $\gamma_1 = 13.994\:301(11)$ {GHz/T}. This corresponds to an optical magnetic field measurement of better than 1 ppm accuracy. From this value we can calculate the fine structure Landé g-factor $g_J\left(6^2P_{3/2}\right) =  1.334\:087\:49(52)$. This is consistent with the previous best measurement, and improves the accuracy by more than two orders of magnitude. We have also measured, for the first time ever, the quadratic diamagnetic shift as $\gamma_2 = 0.4644(35)\:\mathrm{MHz/T^2}$. Our work opens up the field of accurate high field optical magnetometry using atomic cesium. These high accuracy measurements also allow for testing of advanced atomic structure models, as our results are incompatible with the Russel-Saunders coupling value, and the hydrogen-constant-core-model value, by 31 and 7 standard deviations respectively.
\end{abstract}

\maketitle

\section{Introduction}
The field of optical magnetometry has undergone rapid
development during the last couple of decades \cite{Budker2007}. While devices for measuring tiny magnetic fields have matured to the point of emerging practical applications, e.g.\ magnetocardiography and magnetoencephalography \cite{Lembke2014,Hill2020,Sutter2020,Limes2020}, optical magnetometers for high magnetic fields are still at a less advanced level.

Low field optical magnetometry typically works by optical detection of the Larmor precession of optically pumped atomic spins \cite{Budker2007}. High field optical magnetometers, on the other hand, usually rely on measuring the Zeeman shift of the optical absorption lines. Much of the work is focused on the D lines of alkali vapors \cite{George2017,Ciampini2017,Keaveney2019,Klinger2020}. In other work, not directly aimed at magnetometry applications, (nonlinear) spectroscopy and optical pumping of alkali atoms in strong magnetic fields have been studied \cite{Hori1982,Olsen2011,Weller2012,Sargsyan2015,Sargsyan2017,Keaveney2018}.

Accurate measurements of magnetic fields in the tesla range are, currently, typically performed using nuclear magnetic resonance (NMR) spectroscopy on protons in water \cite{Phillips1977}. NMR measurements are highly sensitive, but require application and detection of radio frequency (RF) magnetic pulses. Optical magnetometry provides a completely different way of measuring high magnetic fields, with advantages such as continuous, fast readout, and all optical detection - i.e.\ no metallic or electronic components, and without any applied RF fields. Optical magnetometry also allows for remote detection, e.g.\ measurements on exploding wires \cite{Garn1966,Banasek2016a,Banasek2016b,Gomez2014} and sun spots \cite{Borrero2011}. Indeed the observation of sodium line splitting in sun spots date all the way back to 1870 \cite{Iniesta1996}.

On Earth, high magnetic fields are found in magnetic resonance imaging (MRI) scanners, NMR spectrometers, particle accelerators, fusion reactors, and a range of advanced physics experiments \cite{Battesti2018}.

As is pointed out in \cite{Ciampini2017,Battesti2018} the excited state Landé g-factor determination is limiting the accuracy of rubidium magnetometers. The excited state g-factor for the alkali D lines are all known to about the same accuracy \cite{Arimondo1977}, so this will similarly be limiting the accuracy a cesium magnetometer. In order to enable accurate high field optical magnetometry we here present an improved measurement of the excited state Landé g-factor for the cesium D\textsubscript{2} line, along with the first ever measurement of the diamagnetic shift of this line. By using a 3 T and a 7 T MRI scanner we have very stable and homogeneous high magnetic fields, and at the same time all the hardware needed to accurately determine these fields using NMR spectroscopy. We have realized saturated absorption spectroscopy inside the MRI scanners, in order to surpass the limit of the Doppler broadening. These two advantages combined makes our measurements more than two orders of magnitude more accurate than previous work.

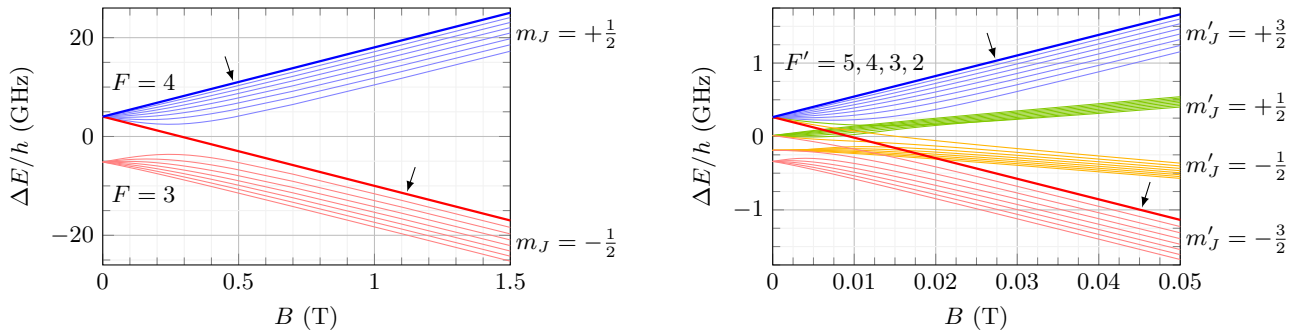
\begin{figure*}[t]
  \centering
  \subfloat[Ground state, $6^2S_{1/2}$, splitting as a function of applied magnetic field. The bold and strongly colored blue line is the $\ket{F,m_F} = \ket{4,4}$ state, and the bold and strongly colored red line is the $\ket{4,-4}$ state. Both are marked with arrows.]{\begin{tikzpicture}
\begin{axis}[
			height              = 5 cm,
			width               = 7 cm,
			xlabel              = $B$ (T),
			ylabel              = $\Delta E/h$ (GHz),
			xmin                = 0,
			xmax                = 1.5,
			ymin                = -26,
			ymax                = 26,
			grid                = both,
			grid style          = {line width=.4pt, draw=gray!10},
            major grid style    = {line width=.4pt, draw=gray!50},
			minor x tick num    = 4,
			minor y tick num    = 3,
			clip                = false]
\ifNotDraft
\addplot [color=my_light_blue] table[x=B,y=2] {Cs_S_1_2.txt};
\addplot [color=my_light_blue] table[x=B,y=3] {Cs_S_1_2.txt};
\addplot [color=my_light_blue] table[x=B,y=4] {Cs_S_1_2.txt};
\addplot [color=my_light_blue] table[x=B,y=5] {Cs_S_1_2.txt};
\addplot [color=my_light_blue] table[x=B,y=6] {Cs_S_1_2.txt};
\addplot [color=my_light_blue] table[x=B,y=7] {Cs_S_1_2.txt};
\addplot [color=my_light_blue] table[x=B,y=8] {Cs_S_1_2.txt};

\addplot [color=my_light_red] table[x=B,y=10] {Cs_S_1_2.txt};
\addplot [color=my_light_red] table[x=B,y=11] {Cs_S_1_2.txt};
\addplot [color=my_light_red] table[x=B,y=12] {Cs_S_1_2.txt};
\addplot [color=my_light_red] table[x=B,y=13] {Cs_S_1_2.txt};
\addplot [color=my_light_red] table[x=B,y=14] {Cs_S_1_2.txt};
\addplot [color=my_light_red] table[x=B,y=15] {Cs_S_1_2.txt};
\addplot [color=my_light_red] table[x=B,y=16] {Cs_S_1_2.txt};

\addplot [color=red, line width = 0.3 mm] table[x=B,y=9] {Cs_S_1_2.txt};
\addplot [color=blue, line width = 0.3 mm] table[x=B,y=1] {Cs_S_1_2.txt};
\fi

\draw [-latex](0.45,16.2) -- (0.45+0.03,16.2-5);
\draw [-latex](1.15,-6.2) -- (1.15-0.03,-6.2-5);

\node at (0.15,11) {$F=4$};
\node at (0.15,-11.7) {$F=3$};
\node at (1.71,20.8) {$m_J=+\frac{1}{2}$};
\node at (1.71,-21.6) {$m_J=-\frac{1}{2}$};
\end{axis}
\end{tikzpicture}
\label{fig:Ground_state_splitting}}
\qquad
  \subfloat[Excited state, $6^2P_{3/2}$, splitting as a function of applied magnetic field. The bold and strongly colored blue line is the $\ket{F',m_F'} = \ket{5,5}$ state, and the bold and strongly colored red line is the $\ket{5,-5}$ state. Both are marked with arrows.]{\begin{tikzpicture}
\begin{axis}[
			height              = 5 cm,
			width               = 7 cm,
			xlabel              = $B$ (T),
			ylabel              = $\Delta E/h$ (GHz),
			xmin                = 0,
			xmax                = 0.05,
			xtick               = {0,0.01,0.02,0.03,0.04,0.05},
			xticklabels         = {0,0.01,0.02,0.03,0.04,0.05},
			scaled x ticks      = false,
			ymin                = -1.75,
			ymax                = 1.75,
			grid                = both,
			grid style          = {line width=.4pt, draw=gray!10},
            major grid style    = {line width=.4pt, draw=gray!50},
			minor tick num      = 3,
			clip                = false]
\ifNotDraft
\addplot [color=my_light_blue] table[x=B,y=2] {Cs_P_3_2.txt};
\addplot [color=my_light_blue] table[x=B,y=3] {Cs_P_3_2.txt};
\addplot [color=my_light_blue] table[x=B,y=4] {Cs_P_3_2.txt};
\addplot [color=my_light_blue] table[x=B,y=5] {Cs_P_3_2.txt};
\addplot [color=my_light_blue] table[x=B,y=6] {Cs_P_3_2.txt};
\addplot [color=my_light_blue] table[x=B,y=7] {Cs_P_3_2.txt};
\addplot [color=my_light_blue] table[x=B,y=8] {Cs_P_3_2.txt};

\addplot [color=my_green] table[x=B,y=9] {Cs_P_3_2.txt};
\addplot [color=my_green] table[x=B,y=10] {Cs_P_3_2.txt};
\addplot [color=my_green] table[x=B,y=11] {Cs_P_3_2.txt};
\addplot [color=my_green] table[x=B,y=12] {Cs_P_3_2.txt};
\addplot [color=my_green] table[x=B,y=13] {Cs_P_3_2.txt};
\addplot [color=my_green] table[x=B,y=14] {Cs_P_3_2.txt};
\addplot [color=my_green] table[x=B,y=15] {Cs_P_3_2.txt};
\addplot [color=my_green] table[x=B,y=16] {Cs_P_3_2.txt};

\addplot [color=my_yellow] table[x=B,y=17] {Cs_P_3_2.txt};
\addplot [color=my_yellow] table[x=B,y=18] {Cs_P_3_2.txt};
\addplot [color=my_yellow] table[x=B,y=19] {Cs_P_3_2.txt};
\addplot [color=my_yellow] table[x=B,y=20] {Cs_P_3_2.txt};
\addplot [color=my_yellow] table[x=B,y=21] {Cs_P_3_2.txt};
\addplot [color=my_yellow] table[x=B,y=22] {Cs_P_3_2.txt};
\addplot [color=my_yellow] table[x=B,y=23] {Cs_P_3_2.txt};
\addplot [color=my_yellow] table[x=B,y=24] {Cs_P_3_2.txt};

\addplot [color=my_light_red] table[x=B,y=26] {Cs_P_3_2.txt};
\addplot [color=my_light_red] table[x=B,y=27] {Cs_P_3_2.txt};
\addplot [color=my_light_red] table[x=B,y=28] {Cs_P_3_2.txt};
\addplot [color=my_light_red] table[x=B,y=29] {Cs_P_3_2.txt};
\addplot [color=my_light_red] table[x=B,y=30] {Cs_P_3_2.txt};
\addplot [color=my_light_red] table[x=B,y=31] {Cs_P_3_2.txt};
\addplot [color=my_light_red] table[x=B,y=32] {Cs_P_3_2.txt};

\addplot [color=red, line width = 0.3 mm] table[x=B,y=25] {Cs_P_3_2.txt};
\addplot [color=blue, line width = 0.3 mm] table[x=B,y=1] {Cs_P_3_2.txt};
\fi

\draw [-latex](0.0263,1.41) -- (0.0263+0.001,1.41-0.35);
\draw [-latex](0.0463,-0.62) -- (0.0463-0.001,-0.62-0.35);

\node at (1e-2,1.0) {$F'=5,4,3,2$};

\node at (5.7e-2,1.4) {$m_J'=+\frac{3}{2}$};
\node at (5.7e-2,0.4) {$m_J'=+\frac{1}{2}$};
\node at (5.7e-2,-0.4) {$m_J'=-\frac{1}{2}$};
\node at (5.7e-2,-1.4) {$m_J'=-\frac{3}{2}$};
\end{axis}
\end{tikzpicture}
\label{fig:Excited_state_splitting}}
\caption{Energy splitting for the ground and excited state, as a function of applied magnetic field.}
\end{figure*}

The data that is presented here is accurate to a level beyond published theoretical values from atomic structure calculations. Hence as a spin-off of our work on magnetometry, we provide data that can be used to test more advanced atomic structure models.

\section{Splitting of the D\textsubscript{2} line}
The transition from the cesium ground state, $6^2S_{1/2}$, to the excited state, $6^2P_{3/2}$, known as the D\textsubscript{2} line is the transition considered in this work. With saturated absorption spectroscopy the hyperfine splitting of this line can readily be resolved \cite{Schmidt1994}. In the following we will review how the line splitting depends on an applied magnetic field in the tesla range. We will consider the Zeeman interaction in detail, and simply take the hyperfine shift to be
\begin{align}
    \Delta E_\mathrm{HFS} = \frac{1}{2} A_\mathrm{HFS} \left(F(F+1)-I(I+1)-J(J+1)\right),
\end{align}
as our results are not sensitive to the details of the hyperfine interaction. For a more detailed treatment of the hyperfine shift including electric quadrupole and magnetic octupole interactions see \cite{Arimondo1977,Arimondo2016,Steck2019}.
\subsection{Zeeman Shift}
As described in \cite{Arimondo1977,Arimondo2016,Steck2019}, the Zeeman shift of a state can be written as
\begin{align}
    \Delta E_\mathrm{Z} = (g_S m_S + g_L m_L + g_I m_I) \mu_\mathrm{B} B.
    \label{eq:kilotesla_fields}
\end{align}
Here $g_S$ is the electron g-factor, $g_L$ is the electron orbital g-factor, and $g_I$ is the nuclear g-factor. We include a finite nuclear mass correction for the orbital g-factor $g_L=m_\mathrm{N}/(m_\mathrm{N}+m_\mathrm{e})\approx1-m_\mathrm{e}/m_\mathrm{N}$ \cite{Arimondo2016,Steck2019,BetheSalpeter1957}. Here $m_\mathrm{e}$ is the electron mass, and $m_\mathrm{N}$ is the nuclear mass. \cref{eq:kilotesla_fields} is the appropriate equation to use in the case of kilotesla fields, i.e.\ in the fine Paschen-Back regime, where the Zeeman shift is large compared to the fine structure. In the case of a Zeeman shift small compared to the fine structure, but large compared to the hyperfine structure, i.e.\ in the hyperfine Paschen-Back regime, we can write the shift as
\begin{align}
    \Delta E_\mathrm{Z} = (g_J m_J + g_I m_I)\mu_\mathrm{B}  B,
\end{align}
with Landé g-factor, $g_J$, approximately given by the Russell-Saunders (RS) coupling value \cite{BetheSalpeter1957,Arimondo2016,Steck2019}
\begin{align}
    g_J =&  g_L \frac{J(J+1)-S(S+1)+L(L+1)}{2J(J+1)} \nonumber \\
         &+ g_S \frac{J(J+1)+S(S+1)-L(L+1)}{2J(J+1)}.
\label{eq:Russel-Saunders}
\end{align}
When the Zeeman shift is small compared to the hyperfine structure, i.e.\ in the Zeeman regime, we can write the shift as
\begin{align}
    \Delta E_\mathrm{Z} = g_F m_F\mu_\mathrm{B}  B,
\end{align}
with Landé g-factor, $g_F$, as given in \cite{Steck2019}.

Numerically diagonalizing the total Hamiltonian composed of the hyperfine Hamiltonian and the Zeeman Hamiltonian (ignoring the small nuclear Zeeman interaction), we can visually inspect the magnetic field dependence of the states, as we transition from the Zeeman regime, into the hyperfine Paschen-Back regime. This is shown for the ground and excited state in \cref{fig:Ground_state_splitting,fig:Excited_state_splitting}, respectively. Notice how the Zeeman interaction acts as a perturbation to the hyperfine splitting at low fields, breaking the degeneracy of the different $m_F$ states. At high fields on the other hand the hyperfine interaction acts as a perturbation to the Zeeman splitting breaking the degeneracy of the different $m_I$ states.

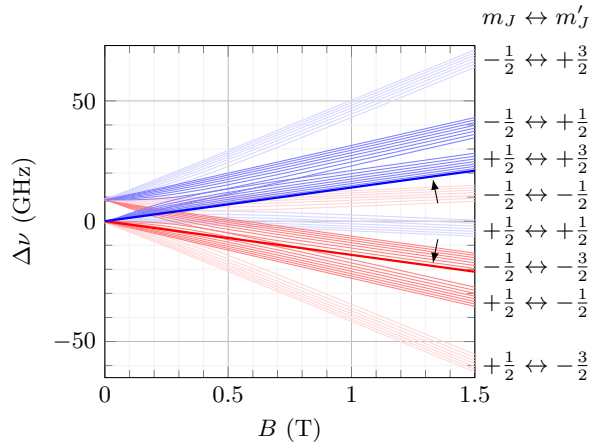
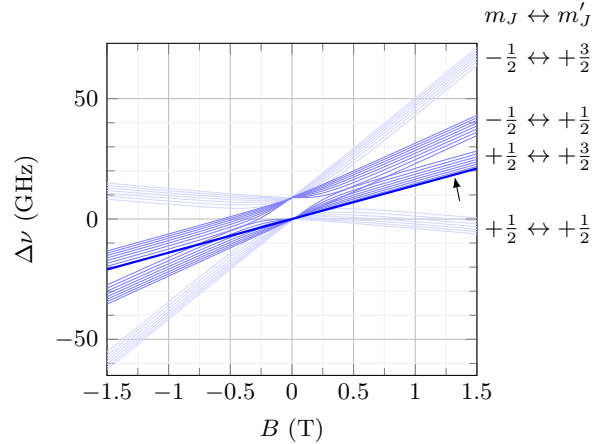
\begin{figure*}[htb]
  \centering
  \subfloat[Splitting of the D\textsubscript{2} line in a magnetic field. The blue lines correspond to $\sigma_+$ transitions, the red lines correspond to $\sigma_-$ transitions. The bold and strongly colored blue line is the transition from the ground state $\ket{4,4}$ to the excited state $\ket{5,5}$. The bold and strongly colored red line is the transition from the ground state $\ket{4,-4}$ to the excited state $\ket{5,-5}$. Both are marked with arrows. Lines in light colors correspond to the weak transitions with $\Delta m_I = \pm1$.]{
\begin{tikzpicture}
\begin{axis}[
			height              = 6 cm,
			width               = 6.5 cm,
			xlabel              = $B$ (T),
			ylabel              = $\Delta\nu$ (GHz),
			xmin                = 0,
			xmax                = 1.5,
			ymin                = -65,
			ymax                = 73,
			grid                = both,
			grid style          = {line width=.4pt, draw=gray!10},
            major grid style    = {line width=.4pt, draw=gray!50},
			minor tick num      = 4,
			clip                = false]
\ifNotDraft
\addplot [color=my_lighter_blue] table[x=B,y=1] {Lines.txt};
\addplot [color=my_lighter_blue] table[x=B,y=2] {Lines.txt};
\addplot [color=my_lighter_blue] table[x=B,y=3] {Lines.txt};
\addplot [color=my_lighter_blue] table[x=B,y=4] {Lines.txt};
\addplot [color=my_lighter_blue] table[x=B,y=5] {Lines.txt};
\addplot [color=my_lighter_blue] table[x=B,y=6] {Lines.txt};
\addplot [color=my_lighter_blue] table[x=B,y=7] {Lines.txt};

\addplot [color=my_lighter_blue] table[x=B,y=24] {Lines.txt};
\addplot [color=my_lighter_blue] table[x=B,y=25] {Lines.txt};
\addplot [color=my_lighter_blue] table[x=B,y=26] {Lines.txt};
\addplot [color=my_lighter_blue] table[x=B,y=27] {Lines.txt};
\addplot [color=my_lighter_blue] table[x=B,y=28] {Lines.txt};
\addplot [color=my_lighter_blue] table[x=B,y=29] {Lines.txt};
\addplot [color=my_lighter_blue] table[x=B,y=30] {Lines.txt};

\addplot [color=my_lighter_red] table[x=B,y=31] {Lines.txt};
\addplot [color=my_lighter_red] table[x=B,y=32] {Lines.txt};
\addplot [color=my_lighter_red] table[x=B,y=33] {Lines.txt};
\addplot [color=my_lighter_red] table[x=B,y=34] {Lines.txt};
\addplot [color=my_lighter_red] table[x=B,y=35] {Lines.txt};
\addplot [color=my_lighter_red] table[x=B,y=36] {Lines.txt};
\addplot [color=my_lighter_red] table[x=B,y=37] {Lines.txt};

\addplot [color=my_lighter_red] table[x=B,y=54] {Lines.txt};
\addplot [color=my_lighter_red] table[x=B,y=55] {Lines.txt};
\addplot [color=my_lighter_red] table[x=B,y=56] {Lines.txt};
\addplot [color=my_lighter_red] table[x=B,y=57] {Lines.txt};
\addplot [color=my_lighter_red] table[x=B,y=58] {Lines.txt};
\addplot [color=my_lighter_red] table[x=B,y=59] {Lines.txt};
\addplot [color=my_lighter_red] table[x=B,y=60] {Lines.txt};

\addplot [color=my_light_red] table[x=B,y=38] {Lines.txt};
\addplot [color=my_light_red] table[x=B,y=39] {Lines.txt};
\addplot [color=my_light_red] table[x=B,y=40] {Lines.txt};
\addplot [color=my_light_red] table[x=B,y=41] {Lines.txt};
\addplot [color=my_light_red] table[x=B,y=42] {Lines.txt};
\addplot [color=my_light_red] table[x=B,y=43] {Lines.txt};
\addplot [color=my_light_red] table[x=B,y=44] {Lines.txt};

\addplot [color=my_light_red] table[x=B,y=46] {Lines.txt};
\addplot [color=my_light_red] table[x=B,y=47] {Lines.txt};
\addplot [color=my_light_red] table[x=B,y=48] {Lines.txt};
\addplot [color=my_light_red] table[x=B,y=49] {Lines.txt};
\addplot [color=my_light_red] table[x=B,y=50] {Lines.txt};
\addplot [color=my_light_red] table[x=B,y=51] {Lines.txt};
\addplot [color=my_light_red] table[x=B,y=52] {Lines.txt};
\addplot [color=my_light_red] table[x=B,y=53] {Lines.txt};

\addplot [color=my_light_blue] table[x=B,y=8] {Lines.txt};
\addplot [color=my_light_blue] table[x=B,y=9] {Lines.txt};
\addplot [color=my_light_blue] table[x=B,y=10] {Lines.txt};
\addplot [color=my_light_blue] table[x=B,y=11] {Lines.txt};
\addplot [color=my_light_blue] table[x=B,y=12] {Lines.txt};
\addplot [color=my_light_blue] table[x=B,y=13] {Lines.txt};
\addplot [color=my_light_blue] table[x=B,y=14] {Lines.txt};
\addplot [color=my_light_blue] table[x=B,y=15] {Lines.txt};

\addplot [color=my_light_blue] table[x=B,y=16] {Lines.txt};
\addplot [color=my_light_blue] table[x=B,y=17] {Lines.txt};
\addplot [color=my_light_blue] table[x=B,y=18] {Lines.txt};
\addplot [color=my_light_blue] table[x=B,y=19] {Lines.txt};
\addplot [color=my_light_blue] table[x=B,y=20] {Lines.txt};
\addplot [color=my_light_blue] table[x=B,y=21] {Lines.txt};
\addplot [color=my_light_blue] table[x=B,y=22] {Lines.txt};

\addplot [color=red, line width = 0.3 mm]  table[x=B,y=45] {Lines.txt};
\addplot [color=blue, line width = 0.3 mm] table[x=B,y=23] {Lines.txt};
\fi
\draw [-latex](1.35,7.5) -- (1.35-0.02,7.5+10);
\draw [-latex](1.35,-7.5) -- (1.35-0.02,-7.5-10);

\node at (1.75,85) {$m_J \leftrightarrow m_J'$};
\node at (1.75,67) {$-\frac{1}{2} \leftrightarrow +\frac{3}{2}$};
\node at (1.75,41) {$-\frac{1}{2} \leftrightarrow +\frac{1}{2}$};
\node at (1.75,26) {$+\frac{1}{2} \leftrightarrow +\frac{3}{2}$};
\node at (1.75,11) {$-\frac{1}{2} \leftrightarrow -\frac{1}{2}$};
\node at (1.75,-4) {$+\frac{1}{2} \leftrightarrow +\frac{1}{2}$};
\node at (1.75,-19) {$-\frac{1}{2} \leftrightarrow -\frac{3}{2}$};
\node at (1.75,-34) {$+\frac{1}{2} \leftrightarrow -\frac{1}{2}$};
\node at (1.75,-60) {$+\frac{1}{2} \leftrightarrow -\frac{3}{2}$};

\end{axis}
\end{tikzpicture}
\label{fig:Lines_splitting}}
\qquad
  \subfloat[Splitting of the D\textsubscript{2} line in a magnetic field extending to negative field values. We only plot $\sigma_+$ transitions. The bold and strongly colored blue line is the transition from the ground state $\ket{4,4}$ to the excited state $\ket{5,5}$. The line is marked with an arrow. Notice how the lines at negative field strengths are mirror images of the $\sigma_-$ transitions in \cref{fig:Lines_splitting}. Lines in light blue correspond to the weak transitions with $\Delta m_I = +1$.]{
\begin{tikzpicture}
\begin{axis}[
			height              = 6 cm,
			width               = 6.5 cm,
			xlabel              = $B$ (T),
			ylabel              = $\Delta\nu$ (GHz),
			xmin                = -1.5,
			xmax                = 1.5,
			ymin                = -65,
			ymax                = 73,
			grid                = both,
			grid style          = {line width=.4pt, draw=gray!10},
            major grid style    = {line width=.4pt, draw=gray!50},
			minor x tick num    = 1,
			minor y tick num    = 4,
			xtick               = {-1.5,-1.0,-0.5,0.0,0.5,1.0,1.5},
			clip                = false]
\ifNotDraft
\addplot [color=my_lighter_blue] table[x=B,y=1] {Lines.txt};
\addplot [color=my_lighter_blue] table[x=B,y=2] {Lines.txt};
\addplot [color=my_lighter_blue] table[x=B,y=3] {Lines.txt};
\addplot [color=my_lighter_blue] table[x=B,y=4] {Lines.txt};
\addplot [color=my_lighter_blue] table[x=B,y=5] {Lines.txt};
\addplot [color=my_lighter_blue] table[x=B,y=6] {Lines.txt};
\addplot [color=my_lighter_blue] table[x=B,y=7] {Lines.txt};

\addplot [color=my_lighter_blue] table[x=B,y=24] {Lines.txt};
\addplot [color=my_lighter_blue] table[x=B,y=25] {Lines.txt};
\addplot [color=my_lighter_blue] table[x=B,y=26] {Lines.txt};
\addplot [color=my_lighter_blue] table[x=B,y=27] {Lines.txt};
\addplot [color=my_lighter_blue] table[x=B,y=28] {Lines.txt};
\addplot [color=my_lighter_blue] table[x=B,y=29] {Lines.txt};
\addplot [color=my_lighter_blue] table[x=B,y=30] {Lines.txt};

\addplot [color=my_lighter_blue] table[x=-B,y=31] {Lines.txt};
\addplot [color=my_lighter_blue] table[x=-B,y=32] {Lines.txt};
\addplot [color=my_lighter_blue] table[x=-B,y=33] {Lines.txt};
\addplot [color=my_lighter_blue] table[x=-B,y=34] {Lines.txt};
\addplot [color=my_lighter_blue] table[x=-B,y=35] {Lines.txt};
\addplot [color=my_lighter_blue] table[x=-B,y=36] {Lines.txt};
\addplot [color=my_lighter_blue] table[x=-B,y=37] {Lines.txt};

\addplot [color=my_lighter_blue] table[x=-B,y=54] {Lines.txt};
\addplot [color=my_lighter_blue] table[x=-B,y=55] {Lines.txt};
\addplot [color=my_lighter_blue] table[x=-B,y=56] {Lines.txt};
\addplot [color=my_lighter_blue] table[x=-B,y=57] {Lines.txt};
\addplot [color=my_lighter_blue] table[x=-B,y=58] {Lines.txt};
\addplot [color=my_lighter_blue] table[x=-B,y=59] {Lines.txt};
\addplot [color=my_lighter_blue] table[x=-B,y=60] {Lines.txt};

\addplot [color=my_light_blue] table[x=B,y=8] {Lines.txt};
\addplot [color=my_light_blue] table[x=B,y=9] {Lines.txt};
\addplot [color=my_light_blue] table[x=B,y=10] {Lines.txt};
\addplot [color=my_light_blue] table[x=B,y=11] {Lines.txt};
\addplot [color=my_light_blue] table[x=B,y=12] {Lines.txt};
\addplot [color=my_light_blue] table[x=B,y=13] {Lines.txt};
\addplot [color=my_light_blue] table[x=B,y=14] {Lines.txt};
\addplot [color=my_light_blue] table[x=B,y=15] {Lines.txt};

\addplot [color=my_light_blue] table[x=B,y=16] {Lines.txt};
\addplot [color=my_light_blue] table[x=B,y=17] {Lines.txt};
\addplot [color=my_light_blue] table[x=B,y=18] {Lines.txt};
\addplot [color=my_light_blue] table[x=B,y=19] {Lines.txt};
\addplot [color=my_light_blue] table[x=B,y=20] {Lines.txt};
\addplot [color=my_light_blue] table[x=B,y=21] {Lines.txt};
\addplot [color=my_light_blue] table[x=B,y=22] {Lines.txt};

\addplot [color=my_light_blue] table[x=-B,y=38] {Lines.txt};
\addplot [color=my_light_blue] table[x=-B,y=39] {Lines.txt};
\addplot [color=my_light_blue] table[x=-B,y=40] {Lines.txt};
\addplot [color=my_light_blue] table[x=-B,y=41] {Lines.txt};
\addplot [color=my_light_blue] table[x=-B,y=42] {Lines.txt};
\addplot [color=my_light_blue] table[x=-B,y=43] {Lines.txt};
\addplot [color=my_light_blue] table[x=-B,y=44] {Lines.txt};

\addplot [color=my_light_blue] table[x=-B,y=46] {Lines.txt};
\addplot [color=my_light_blue] table[x=-B,y=47] {Lines.txt};
\addplot [color=my_light_blue] table[x=-B,y=48] {Lines.txt};
\addplot [color=my_light_blue] table[x=-B,y=49] {Lines.txt};
\addplot [color=my_light_blue] table[x=-B,y=50] {Lines.txt};
\addplot [color=my_light_blue] table[x=-B,y=51] {Lines.txt};
\addplot [color=my_light_blue] table[x=-B,y=52] {Lines.txt};
\addplot [color=my_light_blue] table[x=-B,y=53] {Lines.txt};

\addplot [color=blue, line width = 0.3 mm] table[x=-B,y=45] {Lines.txt};
\addplot [color=blue, line width = 0.3 mm] table[x=B,y=23] {Lines.txt};
\fi

\draw [-latex](1.36,7.5) -- (1.36-0.04,7.5+10);

\node at (2,85) {$m_J \leftrightarrow m_J'$};
\node at (2,67) {$-\frac{1}{2} \leftrightarrow +\frac{3}{2}$};
\node at (2,41) {$-\frac{1}{2} \leftrightarrow +\frac{1}{2}$};
\node at (2,26) {$+\frac{1}{2} \leftrightarrow +\frac{3}{2}$};
\node at (2,-4) {$+\frac{1}{2} \leftrightarrow +\frac{1}{2}$};

\end{axis}
\end{tikzpicture}
\label{fig:Lines_splitting_negative}}
\caption{Splitting of the cesium D\textsubscript{2} line for $\sigma_\pm$ transitions.}
\end{figure*}

\subsection{Extreme Angular Momentum States}
Now we turn our attention to the $\sigma_+$ transition between the extreme angular momentum states, i.e.\ the transition with ground state of maximum total and projected angular momentum, $\ket{F, m_F}=\ket{4,4}$, and excited state of maximum total and projected angular momentum, $\ket{F',m'_F}=\ket{5,5}$. The two states of this transition do not mix with any of the other angular momentum states, when we transition from the Zeeman regime into the hyperfine Paschen-Back regime, and the good quantum numbers change. Even when we enter the fine Paschen-Back regime there is no mixing. The labeling of the states in the different regimes is summarized in \cref{tab:Quantum_Numbers}.
\begin{table}[htb]
\begin{center}
\begin{tabular}{r c c c}
 \hline\hline
&  & Hyperfine & Fine \\
 & Zeeman & Paschen-Back & Paschen-Back \\
 \hline
  & $\ket{F, m_F}$ & $\ket{J, m_J, I, m_I}$ & $\ket{L, m_L, S, m_S, I, m_I}$ \\[0.2ex] 
 \hline\hline
 Ground state & $\ket{4,4}$ & $\ket*{\frac{1}{2},\frac{1}{2}, \frac{7}{2},\frac{7}{2}}$ & $\ket*{0,0,\frac{1}{2},\frac{1}{2},\frac{7}{2},\frac{7}{2}}$ \\ [0.3ex] 
 \hline
 Excited state & $\ket{5,5}$ & $\ket*{\frac{3}{2},\frac{3}{2},\frac{7}{2},\frac{7}{2}}$ & $\ket*{1,1,\frac{1}{2},\frac{1}{2},\frac{7}{2},\frac{7}{2}}$ \\[0.3ex] 
 \hline\hline
\end{tabular}
\end{center}
\caption{Labeling of the extreme angular momentum states in the various regimes.}
\label{tab:Quantum_Numbers}
\end{table}
Notice how the projected angular momentum quantum numbers are equal to the total angular momentum quantum numbers in all regimes, for both states. The frequency shifts of these states are linearly dependent on the magnetic field in all regimes, and also in the intermediate regimes. Hence also the $\sigma_+$ transition between the two states is linearly dependent on the magnetic field.

In \cref{fig:Lines_splitting} the $\sigma_\pm$ transitions in the hyperfine Paschen-Back regime are shown, along with their extrapolations into the Zeeman regime. The electric dipole allowed strong $\sigma_\pm$ transitions are those obeying the selection rules $\Delta m_J = \pm 1$  \cite{Woodgate1980} and, by conservation of angular momentum, $\Delta m_I = 0$. The linearly dependent $\sigma_+$ transition is highlighted in strong blue color. The weaker transitions with $\Delta m_I = \pm 1$, forbidden in the high field limit are also shown.

Highlighted in strong red color in \cref{fig:Lines_splitting} is the $\sigma_-$ transition $\ket{4,-4}\leftrightarrow\ket{5,-5}$, i.e.\ the transition between the negative extreme angular momentum states, which are similarly linearly dependent on the magnetic field. We can view this transition simply as the transition $\ket{4,4}\leftrightarrow\ket{5,5}$ at negative magnetic fields: By reversing the quantization axis, we reverse the magnetic field, the sign of $m_F$, and the handedness of the circularly polarized light. This is shown in \cref{fig:Lines_splitting_negative}, along with the other $\sigma_+$ transitions. An experimental overview confirming this splitting pattern, from $-1.5$ T to $+1.5$ T, is presented in Appendix H.

Taking into account the tiny diamagnetic shift \cite{Garstang1977,Ciampini2017,Huttner1996,Otto2002} of the states, the linear shift of the transition between the extreme angular momentum states is supplemented by a quadratic one (not shown in \cref{fig:Ground_state_splitting,fig:Excited_state_splitting,fig:Lines_splitting,fig:Lines_splitting_negative}). The resulting model constitutes a good operational description of the magnetic field dependence of the line, making it highly useful for accurate magnetometry as described in the next section.

\section{High Field Magnetometry With Cesium}
In the following we will use the quantum numbers of the hyperfine Paschen-Back regime, $\ket{J,m_J,I,m_I}$. Following the conventions of \cite{Arimondo1977,Arimondo2016,Steck2019} for the linear part and the conventions of \cite{Ciampini2017} for the quadratic part, we write the magnetic field dependence of the ground state frequency shift as
\begin{align}
    \Delta\nu_\mathrm{g} = & \left(g_J\left(6^2S_{1/2}\right) m_J + g_I m_I\right)\frac{\mu_\mathrm{B}}{h} B \nonumber\\
                        & + \frac{\xi^\mathrm{dia}_\mathrm{g}}{h} B^2\nonumber\\
                = &\left(g_J\left(6^2S_{1/2}\right) \cdot\sfrac{1}{2} + g_I \cdot\sfrac{7}{2}\right)\frac{\mu_\mathrm{B}}{h} B \nonumber\\
                        & + \frac{\xi^\mathrm{dia}_\mathrm{g}}{h} B^2,
\end{align}
and similarly for the excited state
\begin{align}
    \Delta\nu_\mathrm{e} = &\left(g_J\left(6^2P_{3/2}\right) \cdot\sfrac{3}{2} + g_I \cdot\sfrac{7}{2}\right)\frac{\mu_\mathrm{B}}{h} B \nonumber\\
                        & + \frac{\xi^\mathrm{dia}_\mathrm{e}}{h} B^2.
\end{align}
Notice how the nuclear Zeeman shifts of the two states are the same, so that they do not contribute to the magnetic field dependence of the transition. The magnetic field dependence of the transition becomes
\begin{align}
    \Delta\nu = &\Delta\nu_\mathrm{e} - \Delta\nu_\mathrm{g}\nonumber\\
              = &\left(g_J\left(6^2P_{3/2}\right) \cdot\sfrac{3}{2} - g_J\left(6^2S_{1/2}\right)\cdot\sfrac{1}{2}\right)\frac{\mu_\mathrm{B}}{h} B\nonumber\\
              &+\frac{1}{h}\left(\xi_\mathrm{e}^{\mathrm{dia}}-\xi_\mathrm{g}^{\mathrm{dia}}\right) B^2.
\end{align}
Defining for simplicity
\begin{align}
    \label{eq:def_linear_factor}
    \gamma_1&\equiv \frac{\mu_\mathrm{B}}{h}\left(g_J\left(6^2P_{3/2}\right)\cdot\sfrac{3}{2} - g_J\left(6^2S_{1/2}\right)\cdot\sfrac{1}{2}\right),\\
    \gamma_2&\equiv\frac{1}{h}\left(\xi_\mathrm{e}^{\mathrm{dia}}-\xi_\mathrm{g}^{\mathrm{dia}}\right),
\end{align}
we get
\begin{align}
    \Delta\nu = \gamma_1 B + \gamma_2 B^2.
    \label{eq:Freq_shift}
\end{align}
In practice, for the studied range of magnetic fields it will be useful to parameterize the frequency shift as
\begin{align}
    \Delta\nu = \gamma_0 + \gamma_1\sigma B + \gamma_2\sigma^2 B^2,
    \label{eq:Freq_shift_in_practise}
\end{align}
where $\gamma_0$ is an experimental offset in the frequency shift measurement, and $\sigma$ is a factor describing the field shift introduced by the probe (the structure containing the cesium vapor), such that $B$ is defined as the field in the absence of the probe. Ideally the offset, $\gamma_0$, should be small compared to the linewidth of the transition, and the probe field shift parameter, $\sigma$, should deviate from 1 only by a few ppm.

Using this relation, accurate high field magnetometry can be performed, by measuring the optical frequency shift, $\Delta\nu$. However, knowledge of $\gamma_1$ is limited by the large uncertainty on the excited state Landé g-factor, $g_J\left(6^2P_{3/2}\right)$, and neither $\gamma_2$, nor its constituents, have ever been measured before.

In order to enable accurate high field magnetometry with cesium, we here present a highly improved measurement of $\gamma_1$, and therefore also of $g_J\left(6^2P_{3/2}\right)$. We also present a first ever measurement of $\gamma_2$.

\section{Current Best Numbers}
The value for the cesium ground state Landé g-factor is in \cite{Arimondo1977} determined from experimental data as 2.002 540 32(20). This calculation is based on very precise measurements of the free electron g-factor, $g(e)$, the ratio between the rubidium ground state and the free electron g-factors, $g_J(Rb)/g(e)$, and the ratio between the cesium and the rubidium ground state g-factors, $g_J(Cs)/g_J(Rb)$. As done in \cite{George2017,Ciampini2017} for rubidium, we can redo the calculations with updated values for $g(e)$ \cite{Tiesinga2021}, and $g_J(Rb)/g(e)$ \cite{Tiedeman1977}, to arrive at
\begin{align}
    g_J\left(6^2S_{1/2}\right) = 2.002\:540\:261(27).
    \label{eq:ground_state_g_factor}
\end{align}
This is about an order of magnitude improvement in precision, compared to \cite{Arimondo1977}. This value is the current best estimate on the ground state g-factor.

From \cite{Arimondo1977,Steck2019} we have that the best measurement of the excited state Landé g-factor is
\begin{align}
    g_J\left(6^2P_{3/2}\right) = 1.334\:00(30),
    \label{eq:excited_state_g_factor}
\end{align}
as measured in 1975 by Abele et al. \cite{Abele1975}.

The RS coupling value, \cref{eq:Russel-Saunders}, can be evaluated using either the free electron g-factor for $g_S$, or as suggested in \cite{Ciampini2017}, the ground state g-factor, $g_J\left(6^2S_{1/2}\right)$ in \cref{eq:ground_state_g_factor}, since according to \cref{eq:Russel-Saunders} the two should be identical.
\begin{align}
    g_J\left(6^2P_{3/2}\right)_{g_S = g(e)}                       &= 1.334\:103\:68 \label{eq:g-factor_RS_free},\\
    g_J\left(6^2P_{3/2}\right)_{g_S = g_J\left(6^2S_{1/2}\right)} &= 1.334\:177\:33.
    \label{eq:g-factor_RS_ground}
\end{align}
It should be noted that \cref{eq:excited_state_g_factor} agrees with both \cref{eq:g-factor_RS_free,eq:g-factor_RS_ground}.

Using \cref{eq:ground_state_g_factor,eq:excited_state_g_factor}, and the value in \cite{Tiesinga2021} for $\mu_\mathrm{B}$, in \cref{eq:def_linear_factor} we have that
\begin{align}
    \gamma_1= 13.992\:5(63) \:\mathrm{GHz/T}.
    \label{eq:gamma_1_old}
\end{align}
This is the current best value for $\gamma_1$.

The value of $\xi^\mathrm{dia}$ for these states can be estimated, using the hydrogen-constant-core-model (HCCM) \cite{Garstang1977,Ciampini2017,Huttner1996,Otto2002}, to be 
\begin{align}
    \xi^\mathrm{dia} = \frac{5e^2a_0^2}{8m_\mathrm{e}}\left(1+\frac{1-3l(l+1)}{5(n^*)^2}\right)\nonumber \\
    \times\frac{l(l+1)+m_l^2-1}{(2l-1)(2l+3)}(n^*)^4.
\end{align}
Notice that we follow the conventions of \cite{Ciampini2017} for the diamagnetic shift. Here $e$ is the electron charge, $a_0$ is the Bohr radius, and $m_\mathrm{e}$ is the electron mass \cite{Tiesinga2021}. Using the reduced electron mass, due to the finite nuclear mass, is not relevant for the first five digits. Using the effective principal quantum numbers, $n^*$, of \cite{Lorenzen1984} we get a quadratic shift of 0.3202 MHz/T$^2$ for the ground state and 0.7602 MHz/T$^2$ for the excited state, resulting in an expected diamagnetic shift parameter for the D\textsubscript{2} line
\begin{align}
    \gamma_2 = 0.4400\:\mathrm{MHz/T^2}.
    \label{eq:gamma_2_model}
\end{align}

\section{Method}
\subsection{Realizing Saturated Absorption Spectroscopy Inside an MRI Scanner}
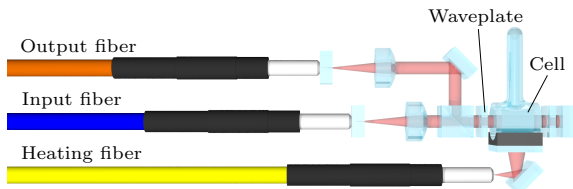
\begin{figure}[ht]
    \centering
    \begin{tikzpicture}
        \node[anchor=south west,inner sep=0] (image) at (0,0) {\includegraphics[width=\linewidth]{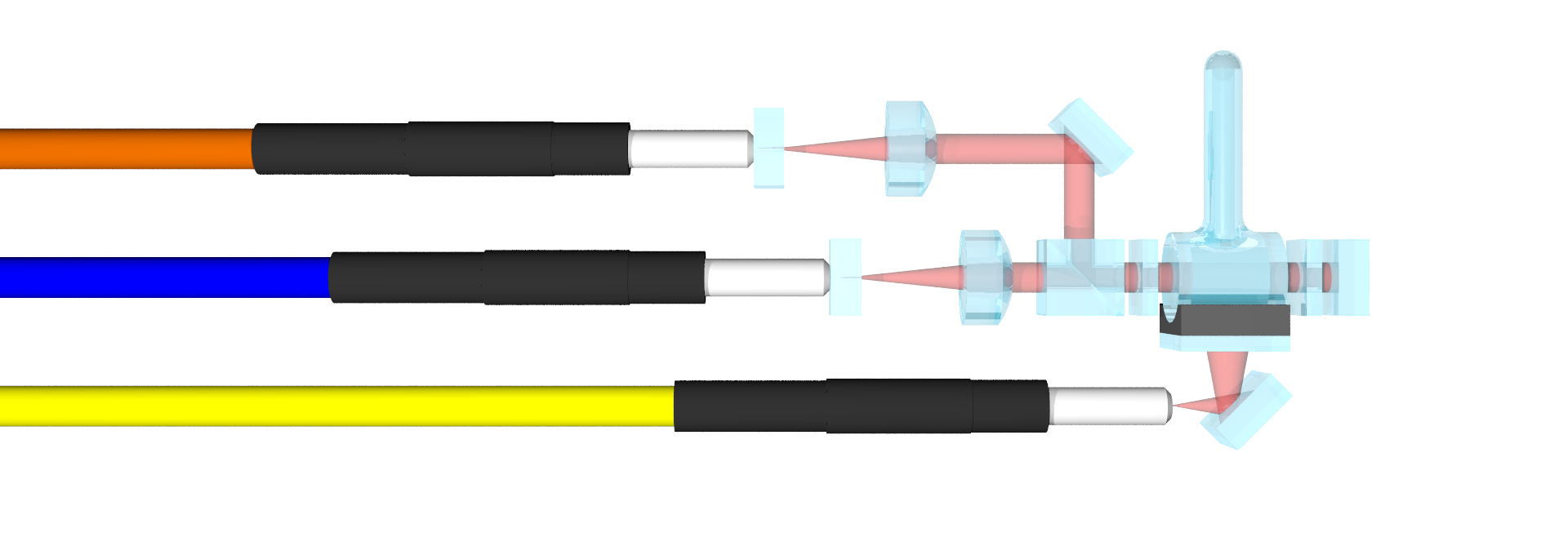}};
        \begin{scope}[x={(image.south east)},y={(image.north west)}]
            \node at (0.11,0.82) {\scriptsize Output fiber};
            \node at (0.1,0.59) {\scriptsize Input fiber};
            \node at (0.115,0.36) {\scriptsize Heating fiber};
            \node at (0.72,0.95) {\scriptsize Waveplate};
            \draw [line width=0.05mm,-](0.724,0.90) -- (0.734,0.54);
            \node at (0.83,0.75) {\scriptsize Cell};
            \draw [line width=0.05mm,-](0.83,0.70) -- (0.80,0.54);
        \end{scope}
    \end{tikzpicture}
    \caption{Optics for performing saturated absorption spectroscopy. Probe light enters through the blue PM fiber, and exits through the orange MM fiber. A high power laser beam delivered through the yellow MM fiber is heating the vapor cell. The angle of the quarter waveplate defines the handedness of the circularly polarized probe light.}
    \label{fig:Probe_a}
\end{figure}

In order to perform saturated absorption spectroscopy inside an MRI scanner we have developed a completely non-metallic fiber-coupled probe containing all the necessary optics shown in \cref{fig:Probe_a}.
The probe light is delivered to the probe in a single mode polarization maintaining (PM) fiber, and returned for detection through a multimode (MM) fiber. To keep the probe and the probed volume small the optical path length in the cesium vapor cell is only 5 mm long, so in order to increase absorption, the cell is heated with a high power laser, delivered through a MM fiber. The fibers are 19 m long.

\begin{figure}[htb]
    \centering
     \begin{tikzpicture}
        \node[anchor=south west,inner sep=0] (image) at (0,0) {\includegraphics[width=\linewidth]{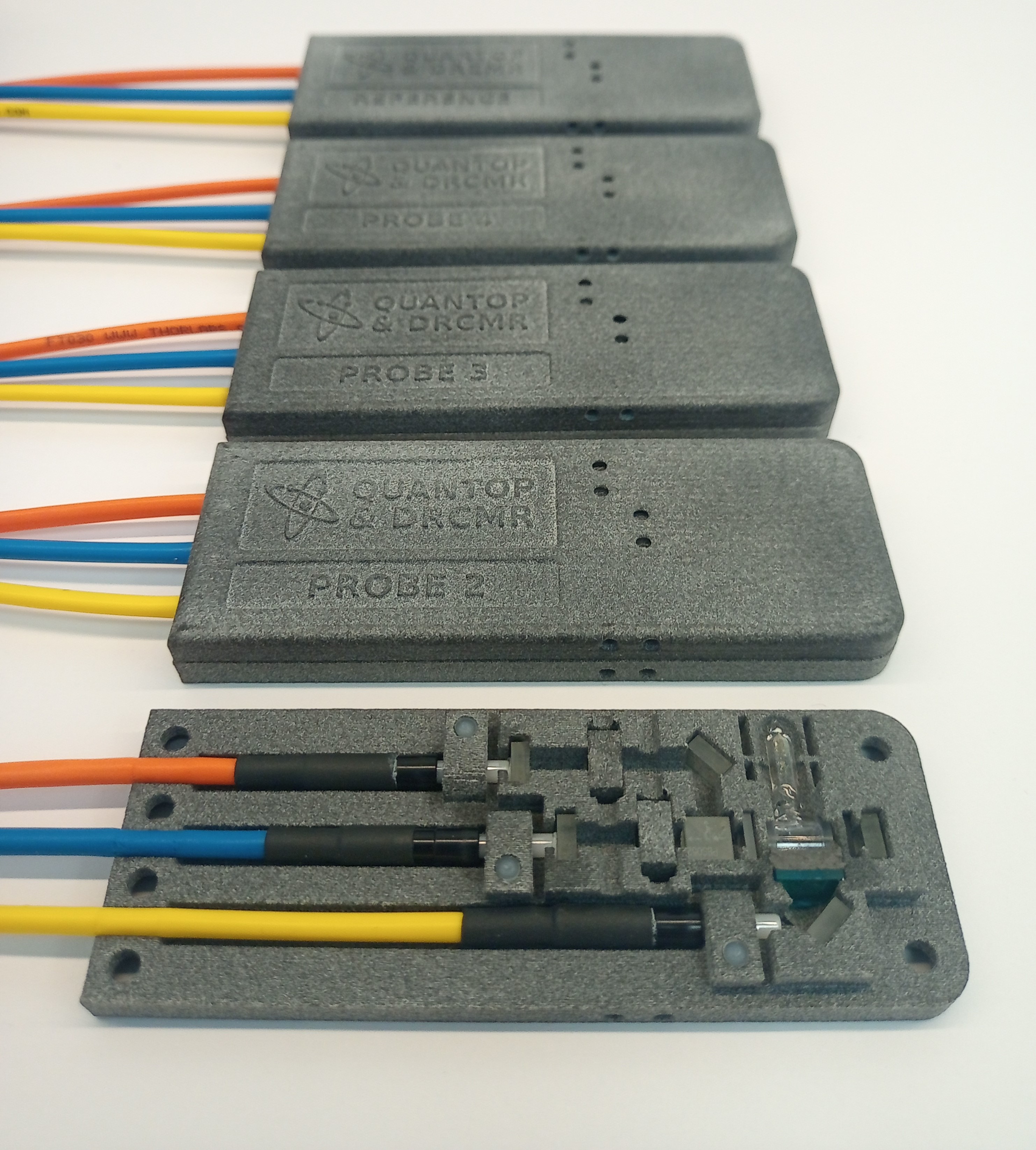}};
        \begin{scope}[x={(image.south east)},y={(image.north west)}]
            \node at (0.5,0.06) {90 mm};
            \draw [line width=0.05mm,-](0.1,0.09) -- (0.9,0.09);
        \end{scope}
    \end{tikzpicture}
    \caption{The physical realisation of the five probes. The cover is removed from probe 1 to show the optics inside.}
    \label{fig:Probes_photo}
\end{figure}

All the optics of the probe is mounted in a $90\times 33\times 10\: \mathrm{mm}^3$ 3D printed nylon enclosure. A total of five probes has been assembled, as shown in \cref{fig:Probes_photo}.
Notice that the quarter waveplate can easily be turned 90\textdegree\ to shift the handedness of the circular polarization. For details on the probe design see Appendix A.

The magnetic susceptibility of the various components that make up the probes has been measured, and the associated field shift at the position of the probing laser beam inside the vapor cell is determined to be
\begin{align}
    \sigma = 1 + 0.93(50)\times 10^{-6}.
    \label{eq:Field_shift_four_probes}
\end{align}
For details on this, see Appendix B.

\subsection{Proton Spectroscopy}
We can very precisely measure the magnetic field inside the MRI scanners (Philips Achieva 3 T and 7 T systems) by proton NMR spectroscopy. Using the hardware of the scanner we excite hydrogen nuclei, in a spherical container of ultra pure water, with an RF pulse and read out the precession frequency, $\nu_\mathrm{p}$, inductively. From this the magnetic field can be then be calculated as
\begin{align}
    B = \frac{\nu_\mathrm{p}}{\gamma'_\mathrm{p}(t)},
    \label{eq:proton_spectroscopy}
\end{align}
where $\gamma'_\mathrm{p}(t)$ is the \textit{shielded proton gyromagnetic ratio} corrected for a small dependence on the temperature, $t$. For further details see Appendix C. We find the field homogeneity to be on the level of 0.3 ppm, over the relevant volume.

\subsection{Sideband Spectroscopy}
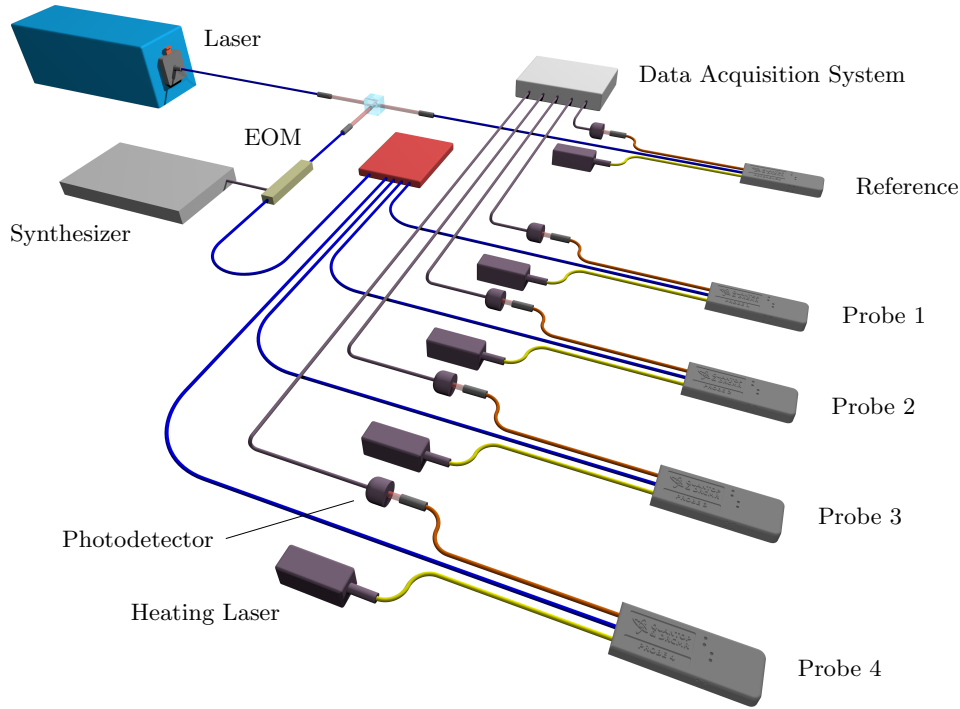
\begin{figure*}[t]
    \centering
    \begin{tikzpicture}
        \node[anchor=south west,inner sep=0] (image) at (0,0) {\includegraphics[width=\textwidth]{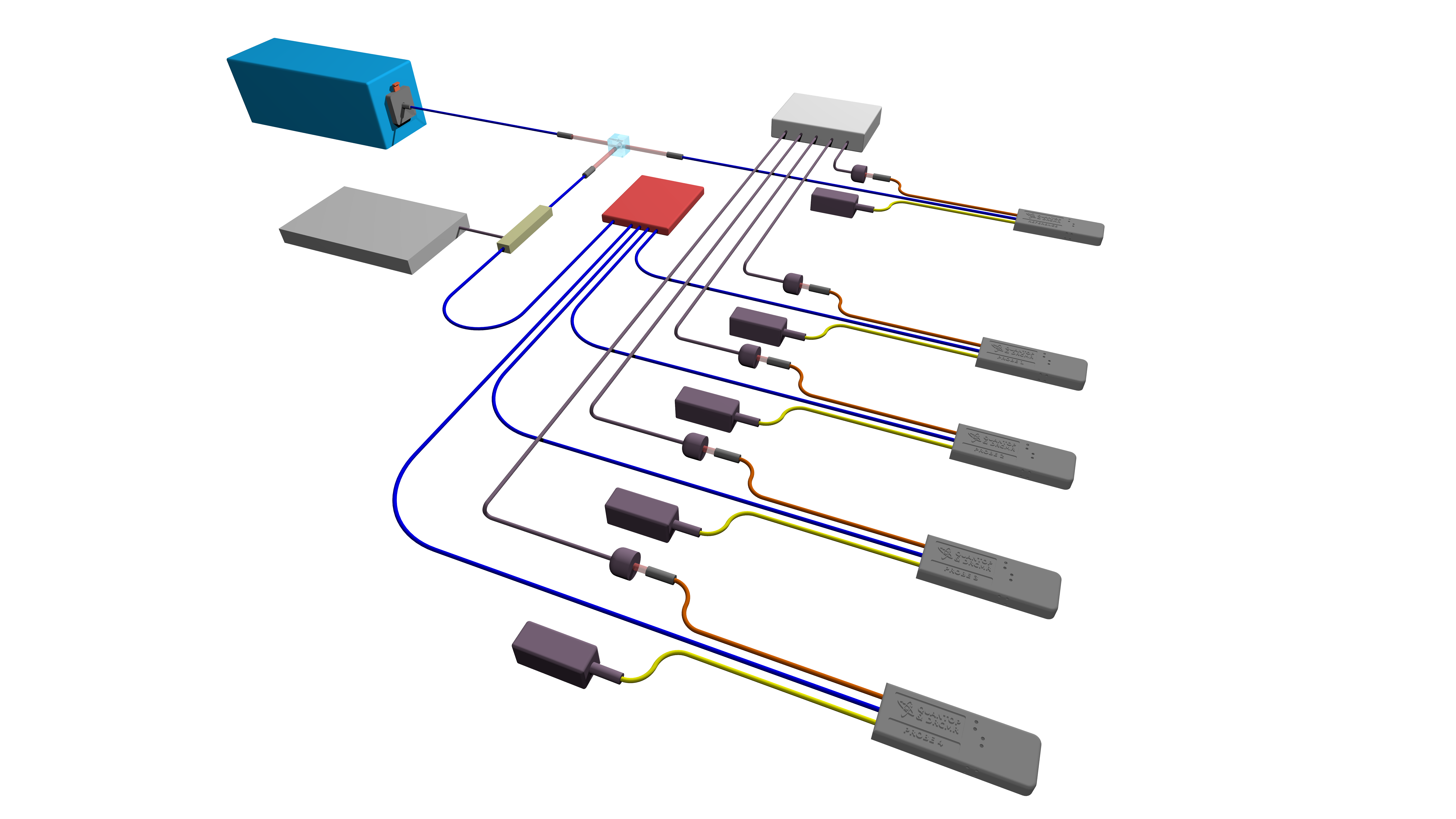}};
        \begin{scope}[x={(image.south east)},y={(image.north west)}]
            \node at (0.32,0.91) {Laser};
            \node at (0.72,0.86) {Data Acquisition System};
            \node at (0.20,0.65) {Synthesizer};
            \node at (0.35,0.78) {EOM};
            \node at (0.30,0.15) {Heating Laser};
            \node at (0.25,0.25) {Photodetector};
            \draw [line width=0.05mm,-](0.315,0.25) -- (0.41,0.30);
            \node at (0.82,0.715) {Reference};
            \node at (0.803,0.545) {Probe 1};
            \node at (0.795,0.425) {Probe 2};
            \node at (0.785,0.28) {Probe 3};
            \node at (0.77,0.08) {Probe 4};
            \end{scope}
    \end{tikzpicture}
    \caption{The optical setup. A laser beam is split in two: One part is send to the reference probe, which is put inside a magnetic shield. The other part is passed through an EOM, and then split into four, and send to each of the four probes, which are located inside the MRI scanner.}
    \label{fig:Setup}
\end{figure*}
Four of the probes are placed in the magnetic field in the center of the MRI scanner, and the fifth is placed in a magnetic shield far away from the MRI scanner. This zero-field reference probe is designed with the quarter-wave plate placed after (instead of before) the cell, such that the probe light polarization is linear. This way, a small residual magnetic field, inside the magnetic shield, only broadens the line instead of shifting it. For the four probes inside the MRI scanner the light is phase-modulated by a high-power, high-frequency electro-optic modulator (EOM), generating multiple strong sidebands. Two of the probes inside the magnetic field are configured with $\sigma_+$ polarization, and two are configured with $\sigma_-$ polarization. This way two measure the field strength as positive, and the other two as negative. Varying the EOM drive frequency we can overlap saturated absorption resonances from carrier (reference cell) and sidebands (probe cells) when scanning the laser frequency and this way measure line shifts as multiples of the EOM drive frequency. The 5\textsuperscript{th} sidebands are used at 7 T, and the 3\textsuperscript{rd} sidebands are used at 3 T. This method drastically reduces sensitivity to nonlinearities of the laser scan. In principle, only two probes are needed inside the MRI scanner, but the redundancy with four probes enables powerful checks for systematic errors. We use a Toptica DL Pro, 852 nm, external cavity diode laser (ECDL), as our probe light source. The setup is shown in \cref{fig:Setup}. For further details on the resonance overlapping method see Appendix D.

\newsavebox\myboxSpectrum
\savebox\myboxSpectrum{ 
\begin{tikzpicture}
\begin{axis}[
            width=4.2cm,
            height=3.3cm,
 			yticklabels             = {},
 			xmin                    = -0.15,
 			xmax                    = 0.05,
 			xtick                   = {-0.15,-0.1,-0.05,0,0.05},
            xticklabels             = {-0.15,-0.1,-0.05,0,0.05},,
 			ymin                    = -0.04,
 			ymax                    = 0.44,
			grid                    = both,
			grid style              = {line width=.4pt, draw=gray!10},
            major grid style        = {line width=.4pt, draw=gray!50},
			minor x tick num        = 1,
			minor y tick num        = 1,
			axis background/.style  = {fill = white},
			xticklabel style        = {font=\scriptsize,yshift=0.2ex},
			clip                    = true]
  \ifNotDraft
  \addplot [color=black] table[x=Freq,y=Ref.]{Broad_spectrum_zoom.txt};
  \addplot [color=blue]  table[x=Freq,y=sigma_+]{Broad_spectrum_zoom.txt};
  \addplot [color=red]  table[x=Freq,y=sigma_-]{Broad_spectrum_zoom.txt};
  \fi
\end{axis}
\end{tikzpicture}
}

\begin{figure*}[htb]
\centering
  \begin{tikzpicture}
  \begin{axis}[
			height              = 5.5 cm,
			width               = \textwidth,
			ylabel              = Transmission (arb. unit),
			xlabel              = $\Delta\nu_\mathrm{L}$ (GHz),
 			yticklabels         = {},
 			xmin                = -0.99612,
 			xmax                = 9.44881,
 			ymin                = -0.6,
 			ymax                = 1.1,
			grid                = both,
			grid style          = {line width=.4pt, draw=gray!10},
            major grid style    = {line width=.4pt, draw=gray!50},
			minor x tick num    = 3,
			minor y tick num    = 2,
			clip                = false]
  \ifNotDraft
  \addplot [color=black] table[x=Freq,y=Ref.]{Broad_spectrum.txt};
  \addplot [color=blue]  table[x=Freq,y=sigma_+]{Broad_spectrum.txt};
  \addplot [color=red]  table[x=Freq,y=sigma_-]{Broad_spectrum.txt};
  \fi
  \node at (axis cs:1.5,-0.55) {\usebox\myboxSpectrum};
  \end{axis}
  \end{tikzpicture}
\caption{A broad scan of the laser frequency, $\Delta\nu_\mathrm{L}$, over the 0 T spectrum (black line), and the $\pm$7 T spectrum as probed by the $\pm$5\textsuperscript{th} sidebands, for \mbox{$\nu_\mathrm{EOM}=19592.24$ MHz} (blue line is $+7$ T and red line is $-7$ T). The frequency axis is calculated by assuming a completely linear scan. The offset between the 0 T resonance and the overlapping $\pm$7 T resonances visible in the inset is due to the diamagnetic shift of about 22 MHz.}
\label{fig:Spectrum_example}
\end{figure*}
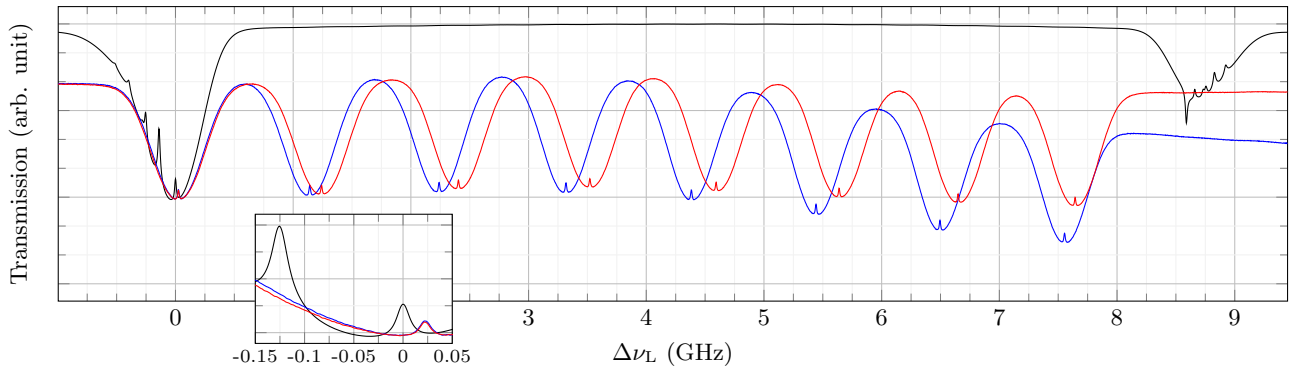

\begin{table*}[htb]
    \centering
    \begin{tabular}{l l r r r r r r r r r}
    \hline\hline
    $B$ & Config. & $t_a$ (\textdegree C) & $t_b$  (\textdegree C) & $\nu_{\mathrm{p},a}$ (Hz) & $\nu_{\mathrm{p},b}$ (Hz) & $\Delta\nu_{\pm B}$ (MHz) & $\Delta\nu_{+B}$ (MHz) & $-\Delta\nu_{-B}$ (MHz) & $\nu_\mathrm{c}$ (Hz) \\
    \hline
    3 T & $\scriptstyle --++$   & 20.3(5) & 20.5(5) & 127 778 093(36)  & 127 778 102(38) & 83\:998.043(88)  & 42\:003.342(88) & 41\:994.668(88) & 9\:999\:991(5) \\
    3 T & $\scriptstyle ++--$   & 20.0(5) & 20.1(5) & 127 777 873(32)  & 127 777 902(41) & 83\:997.954(88)  & 42\:003.311(88) & 41\:994.620(88) & 9\:999\:990(5) \\
    7 T & $\scriptstyle ++--$   & 20.0(5) & 20.0(5) & 298 037 732(60)  & 298 037 734(61) & 195\:922.431(88) & 97\:984.127(88) & 97\:938.342(88) & 9\:999\:992(2) \\
    7 T & $\scriptstyle --++$   & 20.0(5) & 20.0(5) & 298 037 724(42)  & 298 037 723(65) & 195\:922.346(88) & 97\:984.094(88) & 97\:938.225(88) & 9\:999\:992(2) \\
    7 T & $\scriptstyle --++^*$ & 20.0(5) & 19.9(5) & 298 037 732(74)  & 298 037 732(72) & 195\:922.303(88) & 97\:984.058(88) & 97\:938.256(88) & 9\:999\:992(2) \\
    7 T & $\scriptstyle ++--^*$ & 19.4(5) & 19.4(5) & 298 037 716(62)  & 298 037 720(61) & 195\:922.427(88) & 97\:984.168(88) & 97\:938.313(88) & 9\:999\:992(2) \\
    \hline\hline
    \end{tabular}
    \caption{The data points underlying the analysis in this work. The configuration refers to the $\sigma_\pm$ polarization of probe 1, 2, 3, and 4 respectively. The asterisk indicate that the cells in probe 4 and the reference has been interchanged. The temperature, $t$, is measured next to the water reference just prior to the proton spectroscopy. The proton precession frequency, $\nu_\mathrm{p}$, is determined by NMR spectroscopy. Subscripts $a$ and $b$ refer to measurements before and after the optical measurements, respectively. The optical frequency shifts are determined by the overlapping method described in the Appendix D. The ``10 MHz'' MRI scanner clock frequency, $\nu_\mathrm{c}$, also used as the reference the synthesizer driving the EOM, is measured for each data series.}
    \label{tab:final_data}
\end{table*}

\subsection{Data Acquisition Procedure}
A data series is acquired through the following steps:
\begin{itemize}
    \item The water temperature, $t_a$, is noted, and proton NMR spectroscopy is performed on the spherical water container, to measure a precession frequency $\nu_{\mathrm{p},a}$.
    \item The four probes are placed in the MRI scanner, instead of the water container.
    \item The frequency shift between the optical resonance at negative field strength and positive field strength, $\Delta\nu_{\pm B}$, is measured by sideband spectroscopy.
    \item The frequency shift between the optical resonance at zero field and positive field strength, $\Delta\nu_{+B}$, is measured.
    \item The frequency shift between the optical resonance at zero field and negative field strength, $\Delta\nu_{-B}$, is measured.
    \item Finally the water temperature, $t_b$, is noted, and the water container is again placed in the MRI scanner, instead of the probes, to measure a precession frequency $\nu_{\mathrm{p},b}$.
\end{itemize}
The scanner clock, $\nu_\mathrm{c}$, which is also used as the reference clock for the synthesizer driving the EOM, is continuously monitored throughout the experiment, and measured in absolute terms.

A full data acquisition run takes about 45 minutes. For details on how the probes are positioned inside the scanner see Appendix A.

\section{Results}
With the probes positioned in the center of the 7 T MRI scanner, and the EOM driven at \mbox{$\nu_\mathrm{EOM} = $ 19592.24 MHz}, we see a spectrum as in \cref{fig:Spectrum_example}. Notice how the $\sigma_-$ and the $\sigma_+$ transition overlap well with this choice of $\nu_\mathrm{EOM}$. Since it is the 5\textsuperscript{th} sidebands that probe the $\pm$7 T transitions this indicates that the shift from $-7$ T to $+7$ T is about $10\cdot\nu_\mathrm{EOM}$. If no diamagnetic shift existed, these two resonances should also overlap with the 0 T resonance. This is clearly not the case, as they are seen to be about $0.44\:\mathrm{MHz/T^2}\cdot \left(7\:\mathrm{T}\right)^2$ = 22 MHz higher, as predicted by \cref{eq:gamma_2_model}.

To get an accurate, reproducible, and unbiased measurement of the $\nu_\mathrm{EOM}$ that best overlaps the resonances, $\nu_\mathrm{EOM}$ is varied and line positions are fitted. The procedure, considered error sources, and associated uncertainty estimates are described in Appendix D and E.

Six experimental runs are performed: Two at the 3 T scanner - first probe 1 and 2 are configured with $\sigma_-$ polarization, and probe 3 and 4 are configured with $\sigma_+$ polarization - and secondly the opposite configuration is used. Next, the same two configurations are used at the 7 T scanner. Finally the vapor cells in probe 4 and the reference probe are interchanged, and the same two configurations are again used.

For each experimental run a line of data is listed in \cref{tab:final_data}. The data, tabulated chronologically, is acquired over the period from April 17\textsuperscript{th} to May 22\textsuperscript{th} 2022.

\subsection{Calculating the Linear Magnetic Frequency Shift}
By measuring the frequency shift, $\Delta\nu_{\pm B}$, from $-B$ to $+B$, we can eliminate the measurement offset and the quadratic contribution, in \cref{eq:Freq_shift_in_practise}, to find
\begin{align}
    \Delta\nu_{\pm B} = 2\gamma_1\sigma B.
    \label{eq:linear_measurement}
\end{align}
Isolating $\gamma_1$ and using \cref{eq:proton_spectroscopy} we find
\begin{align}
    \gamma_1    &= \frac{\Delta\nu_{\pm B}}{2\sigma B}\\
                &= \frac{\Delta\nu_{\pm B} \cdot\gamma'_\mathrm{p}(t)}{2\sigma\nu_\mathrm{p}}.
    \label{eq:gamma_1_new_calculation}
\end{align}
Notice that this expression contains the ratio of two frequency measurements: the proton precession frequency and the optical frequency shift. This means that it is not sensitive to the absolute accuracy of these frequencies, as long as a single common clock is used, as is the case for our experiment.

For each line in \cref{tab:final_data} two values for $\gamma_1$ are calculated: One based on the optical frequency shift $\Delta\nu_{\pm B}$ (single), and one based on the sum of $\Delta\nu_{+B}$ and $-\Delta\nu_{-B}$ (sum). The proton precession frequency for each line is taken to be the average of the measurement before and after, with the uncertainty taken to be the difference, plus the two individual uncertainties. Similarly for the temperatures, except that we only include the thermometer uncertainty of 0.5 \textdegree C once. The resulting values are displayed in \cref{tab:gamma_1_results}, along with the MRI scanner field strength and the probe polarization configuration. We find the mean of these values to be
\begin{align}
    \gamma_1 = 13.994\:301(11)\:\mathrm{GHz/T}.
    \label{eq:gamma_1_new}
\end{align}
\begin{table}[htb]
    \centering
    \begin{tabular}{r c c c}
    \hline\hline
    $B$ & Config.             & $\gamma_1$ (GHz/T) (single) & $\gamma_1$ (GHz/T) (sum) \\
    \hline
    3 T & $\scriptstyle --++$   & 13.994299(19)       & 13.994294(24)   \\
    3 T & $\scriptstyle ++--$   & 13.994307(20)       & 13.994304(25)   \\
    7 T & $\scriptstyle ++--$   & 13.994304(11)       & 13.994307(13)   \\
    7 T & $\scriptstyle --++$   & 13.994299(11)       & 13.994297(12)   \\
    7 T & $\scriptstyle --++^*$ & 13.994295(12)       & 13.994296(13)   \\
    7 T & $\scriptstyle ++--^*$ & 13.994305(11)       & 13.994309(13)   \\
    \hline\hline
    \end{tabular}
    \caption{The various measurements of $\gamma_1$.}
    \label{tab:gamma_1_results}
\end{table}
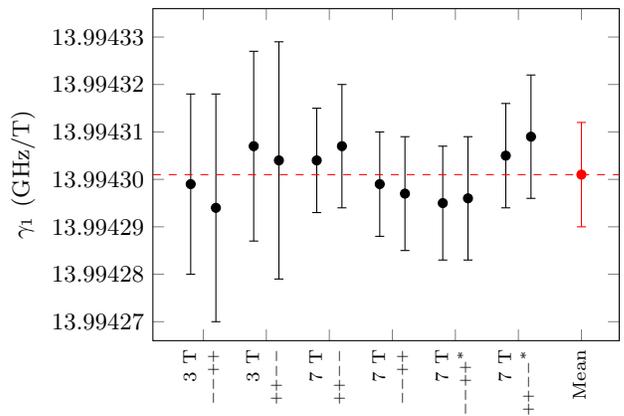
\begin{figure}[htb]
\centering
\begin{tikzpicture}
\begin{axis}[
			height              = 6 cm,
			width               = 0.9\linewidth,
			xmin                = 0.2,
			xmax                = 7.6,
			xtick               = {1,2,3,4,5,6,7},
			xticklabel style    = {rotate=90, 
			                        align=right},
			xticklabels         = {3 T\\$\scriptstyle --++$,3 T\\$\scriptstyle ++--$,7 T\\$\scriptstyle ++--$,7 T\\$\scriptstyle --++$,7 T\\$\scriptstyle --++^*$,7 T\\$\scriptstyle ++--^*$,Mean},
			ymin                = 301-35,
			ymax                = 301+35,
			ytick               = {270,280,290,300,310,320,330},
			yticklabels         = {13.99427,13.99428,13.99429,13.99430,13.99431,13.99432,13.99433},
			ylabel              = $\gamma_1$ (GHz/T),
			grid style          = {line width=.4pt, draw=gray!10},
            major grid style    = {line width=.4pt, draw=gray!50},
			clip                = false,
			xticklabel style    = {font=\scriptsize,yshift=0.2ex}
            ]
\ifNotDraft
\addplot[black,mark=*,mark size = 0.6 mm] plot[ error bars/.cd, y dir=both, y explicit] coordinates {
            (0.8,299) +- (0,19)}; 
\addplot[black,mark=*,mark size = 0.6 mm] plot[ error bars/.cd, y dir=both, y explicit] coordinates {
            (1.2,294) +- (0,24)}; 
\addplot[black,mark=*,mark size = 0.6 mm] plot[ error bars/.cd, y dir=both, y explicit] coordinates {
            (1.8,307) +- (0,20)}; 
\addplot[black,mark=*,mark size = 0.6 mm] plot[ error bars/.cd, y dir=both, y explicit] coordinates {
            (2.2,304) +- (0,25)}; 
\addplot[black,mark=*,mark size = 0.6 mm] plot[ error bars/.cd, y dir=both, y explicit] coordinates {
            (2.8,304) +- (0,11)};  
\addplot[black,mark=*,mark size = 0.6 mm] plot[ error bars/.cd, y dir=both, y explicit] coordinates {
            (3.2,307) +- (0,13)};  
\addplot[black,mark=*,mark size = 0.6 mm] plot[ error bars/.cd, y dir=both, y explicit] coordinates {
            (3.8,299) +- (0,11)};  
\addplot[black,mark=*,mark size = 0.6 mm] plot[ error bars/.cd, y dir=both, y explicit] coordinates {
            (4.2,297) +- (0,12)};  
\addplot[black,mark=*,mark size = 0.6 mm] plot[ error bars/.cd, y dir=both, y explicit] coordinates {
            (4.8,295) +- (0,12)};  
\addplot[black,mark=*,mark size = 0.6 mm] plot[ error bars/.cd, y dir=both, y explicit] coordinates {
            (5.2,296) +- (0,13)};  
\addplot[black,mark=*,mark size = 0.6 mm] plot[ error bars/.cd, y dir=both, y explicit] coordinates {
            (5.8,305) +- (0,11)};  
\addplot[black,mark=*,mark size = 0.6 mm] plot[ error bars/.cd, y dir=both, y explicit] coordinates {
            (6.2,309) +- (0,13)};  
\addplot[red,mark=*,mark size = 0.6 mm] plot[ error bars/.cd, y dir=both, y explicit] coordinates {
            (7,301) +- (0,11)};  
\addplot[red, dashed]
       coordinates {(0.2,301) (7.6,301)};
\fi
\end{axis}
\end{tikzpicture}
\caption{The various values for $\gamma_1$ from \cref{tab:gamma_1_results}, in black, along with the mean value, in red.}
\label{fig:gamma_1_values}
\end{figure}
For the uncertainty we simply take the lowest of the uncertainties from \cref{tab:gamma_1_results}, in recognition that part of the uncertainty comes from the probe field shift, $\sigma$, and is common to all the measurements, just like other systematic error sources also might be at work. The value in \cref{eq:gamma_1_new} represents an improvement in accuracy of more than two orders of magnitude compared to \cref{eq:gamma_1_old}. The data in \cref{tab:gamma_1_results} is shown in \cref{fig:gamma_1_values}.

\subsection{Calculating the Landé g-factor for the Excited State}
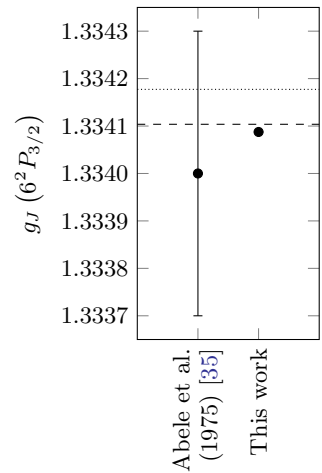
\begin{figure}[htb]
\centering
\begin{tikzpicture}
\begin{axis}[
			height              = 6 cm,
			width               = 4 cm,
			xmin                = 0,
			xmax                = 3,
			xtick               = {1,2},
			xticklabel style    = {align=right},
			xticklabels         = {Abele et al.\\(1975) \cite{Abele1975},This work},
			ymin                = 1.33365,
			ymax                = 1.33435,
			xticklabel style    = {rotate = 90},
			ytick               = {1.3337,1.3338,1.3339,1.3340,1.3341,1.3342,1.3343},
			ylabel              = $g_J\left(6^2P_{3/2}\right)$,
			grid style          = {line width=.4pt, draw=gray!10},
            major grid style    = {line width=.4pt, draw=gray!50},
			clip                = false,
			y tick label style  = { /pgf/number format/.cd,
                                    fixed,
                                    fixed zerofill,
                                    precision=4,
                                    /tikz/.cd }
            ]
\ifNotDraft
\addplot[black,dashed]
    coordinates {(0,1.3341037) (3,1.3341037)};
\addplot[black,densely dotted]
    coordinates {(0,1.3341773) (3,1.3341773)};
\addplot[black,mark=*,mark size = 0.6 mm]
plot[ error bars/.cd,
            y dir=both,
            y explicit
          ] coordinates {
            (1,1.3340) +- (0,0.0003)
            };
\addplot[black,mark=*,mark size = 0.6 mm]
plot[ error bars/.cd,
            y dir=both,
            y explicit,
          ] coordinates {
            (2,1.33408749) +- (0,0.00000052)
            };
\fi
\end{axis}
\end{tikzpicture}
\caption{Comparison of \cref{eq:g-factor_measured} with the best previous measurement, and the two RS values. The dashed line represents the RS value calculated using the free electron g-factor as $g_S$, \cref{eq:g-factor_RS_free}. The dotted line represents the RS value calculated using the ground state g-factor as $g_S$, \cref{eq:g-factor_RS_ground}.}
\label{fig:g_factor_comparison}
\end{figure}

Isolating $g_J\left(6^2P_{3/2}\right)$ in \cref{eq:def_linear_factor} we find
\begin{align}
    g_J\left(6^2P_{3/2}\right) &= \gamma_1\cdot\frac{2h}{3\mu_\mathrm{B}}+\frac{g_J\left(6^2S_{1/2}\right)}{3}\\
    &= 1.334\:087\:49(52).
    \label{eq:g-factor_measured}
\end{align}
This is compared to the previous best value, \cref{eq:excited_state_g_factor}, measured by Abele et al. (1975) \cite{Abele1975}, and the RS values, \cref{eq:g-factor_RS_free,eq:g-factor_RS_ground}, in \cref{fig:g_factor_comparison}.
As with the $\gamma_1$ measurement we see an improvement in accuracy of more than two orders of magnitude. The RS value calculated using the free electron g-factor as $g_S$, \cref{eq:g-factor_RS_free}, is the prediction closest to our result. However the discrepancy is 31 standard deviations, and hence substantial.

\subsection{Determining the Frequency Shift Measurement Offset}
To determine the measurement offset, $\gamma_0$, we reorganize \cref{eq:Freq_shift_in_practise} to read
\begin{align}
    \Delta\nu -\gamma_1\sigma B = \gamma_0 + \gamma_2\sigma^2B^2.
\end{align}
For each line in \cref{tab:final_data} a magnetic field, $B$, is calculated, using \cref{eq:proton_spectroscopy} and the left hand side is calculated for the values $\Delta\nu_{+B}$ and $\Delta\nu_{-B}$, using the new value in \cref{eq:gamma_1_new} for $\gamma_1$.
\begin{figure}[htb]
  \centering
  \begin{tikzpicture}
  \begin{axis}[
			height              = 6 cm,
			width               = 0.9\linewidth,
			xlabel              = $B$ (T),
			ylabel              = $\Delta\nu -\gamma_1\sigma B$ (MHz),
			xmin                = -8,
			xmax                = 8,
			ymin                = 0,
			ymax                = 25,
			grid                = both,
			grid style          = {line width=.4pt, draw=gray!10},
            major grid style    = {line width=.4pt, draw=gray!50},
			minor x tick num    = 1,
			minor y tick num    = 3,
			clip                = true]
  \ifNotDraft
  \addplot [black,only marks,mark size = 0.3 mm] table[x=B_data,y=y_data]{calc_gamma_0_y_data.txt};
  \addplot [color=red]   table[x=B_fit,y=y_fit]{calc_gamma_0_y_fit.txt};
  \fi
  \end{axis}
  \end{tikzpicture}
  \caption{$\gamma_0 + \gamma_2\sigma^2B^2$ data and fit. The least squares fit produces the values $\gamma_0 = 0.159\:\mathrm{MHz}$ and $\gamma_2 = 0.4644\:\mathrm{MHz/T^2}$. This method is only used to estimate $\gamma_0$, and not $\gamma_2$, since it does not provide a good estimate for the uncertainty.}
  \label{fig:quadratic_fit}
\end{figure}
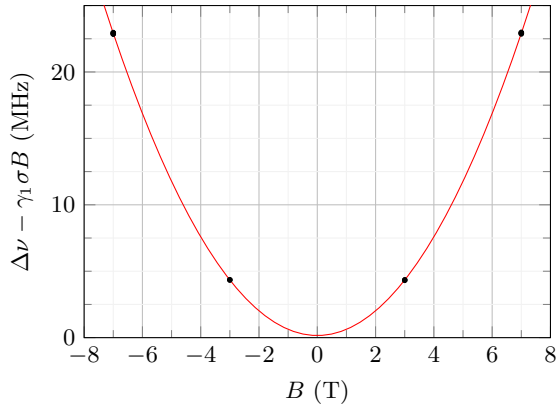
This data set is fitted, without taking uncertainties into account, with the quadratic right hand side, as shown in \cref{fig:quadratic_fit}, to produce the value
\begin{align}
    \gamma_0 = 0.159(159)\:\mathrm{MHz}.
\end{align}
In recognition that this number should be zero, and we do not know exactly the cause of this deviation, we take the uncertainty to be 100 \%, even though the largest residual from the fit is 0.068 MHz. We note that this measurement offset is small compared to the linewidth of the transition, as expected. The clock inaccuracy, listed in \cref{tab:final_data} as $\nu_\mathrm{c}$, is here taken into account, but is found to be negligible.

It should be stressed that $\gamma_0$ is an experimental offset, that depends on the physical implementation of the measurement. We report it here, since it gives a good estimate of the accuracy of the method, and because we need it for the calculations in the next section.

\subsection{Calculating the Quadratic Diamagnetic Shift}
By measuring, and adding, the optical shift from 0 to $+B$, and from 0 to $-B$, we can eliminate the linear part part, in \cref{eq:Freq_shift_in_practise} to get
\begin{align}
    \Delta\nu_{+B} + \Delta\nu_{-B} = 2\gamma_0 + 2\gamma_2\sigma^2B^2.
\end{align}
Isolating $\gamma_2$ and using \cref{eq:proton_spectroscopy} we find
\begin{align}
    \gamma_2&= \frac{\Delta\nu_{+B} + \Delta\nu_{-B} - 2\gamma_0}{2\sigma^2B^2}\\
            &= \frac{(\Delta\nu_{+B} + \Delta\nu_{-B} - 2\gamma_0) \cdot\left(\gamma'_\mathrm{p}(t)\right)^2}{2\sigma^2\nu_\mathrm{p}^2}.
\end{align}
Notice that this expression \textit{is} sensitive to the absolute accuracy of the clock, unlike \cref{eq:gamma_1_new_calculation}. The (in this case, insignificant) frequency correction is made by multiplying the denominator by a factor $\nu_\mathrm{c}\cdot\left(10\:\mathrm{MHz}\right)^{-1}$.

For each line in \cref{tab:final_data} a value for $\gamma_2$ is calculated.
\begin{table}[htb]
    \centering
    \begin{tabular}{r c l c}
    \hline\hline
    $B$ & Config.               & $\gamma_2$ $\left(\mathrm{MHz/T^2}\right)$ & $\gamma_2^\mathrm{d}$ $\left(\mathrm{MHz/T^2}\right)$ \\
    \hline
    3 T & $\scriptstyle --++$   & \:\: 0.4639(190)  & 0.4815(69)     \\
    3 T & $\scriptstyle ++--$   & \:\: 0.4648(190)  & 0.4825(69)     \\
    7 T & $\scriptstyle ++--$   & \:\: 0.4639(35) & 0.4672(13)       \\
    7 T & $\scriptstyle --++$   & \:\: 0.4648(35) & 0.4680(13)       \\
    7 T & $\scriptstyle --++^*$ & \:\: 0.4641(35) & 0.4674(13)       \\
    7 T & $\scriptstyle ++--^*$ & \:\: 0.4647(35) & 0.4679(13)       \\
    \hline\hline
    \end{tabular}
    \caption{The various measurements of $\gamma_2$. Also shown are the values $\gamma_2^\mathrm{d}$, directly calculated without taking the measurement offset, $\gamma_0$, into account.}
    \label{tab:gamma_2_results}
\end{table}
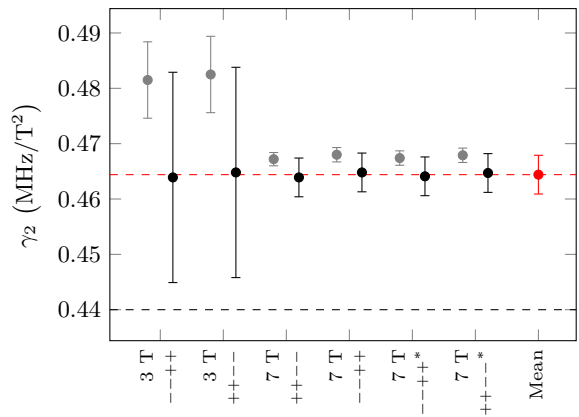
\begin{figure}[htb]
\centering
\begin{tikzpicture}
\begin{axis}[
			height              = 6 cm,
			width               = 0.9\linewidth,
			xmin                = 0.2,
			xmax                = 7.6,
			xtick               = {1,2,3,4,5,6,7},
			xticklabel style    = {rotate=90, 
			                        align=right},
			xticklabels         = {3 T\\$\scriptstyle --++$,3 T\\$\scriptstyle ++--$,7 T\\$\scriptstyle ++--$,7 T\\$\scriptstyle --++$,7 T\\$\scriptstyle --++^*$,7 T\\$\scriptstyle ++--^*$,Mean},
			ymin                = 0.4644-0.03,
			ymax                = 0.4644+0.03,
			ytick               = {0.44,0.45,0.46,0.47,0.48,0.49},
			ylabel              = $\gamma_2$ $\left(\mathrm{MHz/T^2}\right)$,
			grid style          = {line width=.4pt, draw=gray!10},
            major grid style    = {line width=.4pt, draw=gray!50},
			clip                = false,
			xticklabel style    = {font=\scriptsize,yshift=0.2ex}
            ]
\ifNotDraft
\addplot[gray,mark=*,mark size = 0.6 mm] plot[ error bars/.cd, y dir=both, y explicit] coordinates {
            (0.8,0.4815) +- (0,0.0069)}; 
\addplot[black,mark=*,mark size = 0.6 mm] plot[ error bars/.cd, y dir=both, y explicit] coordinates {
            (1.2,0.4639) +- (0,0.0190)}; 
\addplot[gray,mark=*,mark size = 0.6 mm] plot[ error bars/.cd, y dir=both, y explicit] coordinates {
            (1.8,0.4825) +- (0,0.0069)}; 
\addplot[black,mark=*,mark size = 0.6 mm] plot[ error bars/.cd, y dir=both, y explicit] coordinates {
            (2.2,0.4648) +- (0,0.0190)}; 
\addplot[gray,mark=*,mark size = 0.6 mm] plot[ error bars/.cd, y dir=both, y explicit] coordinates {
            (2.8,0.4672) +- (0,0.0012)};  
\addplot[black,mark=*,mark size = 0.6 mm] plot[ error bars/.cd, y dir=both, y explicit] coordinates {
            (3.2,0.4639) +- (0,0.0035)};  
\addplot[gray,mark=*,mark size = 0.6 mm] plot[ error bars/.cd, y dir=both, y explicit] coordinates {
            (3.8,0.4680) +- (0,0.0013)};  
\addplot[black,mark=*,mark size = 0.6 mm] plot[ error bars/.cd, y dir=both, y explicit] coordinates {
            (4.2,0.4648) +- (0,0.0035)};  
\addplot[gray,mark=*,mark size = 0.6 mm] plot[ error bars/.cd, y dir=both, y explicit] coordinates {
            (4.8,0.4674) +- (0,0.0013)};  
\addplot[black,mark=*,mark size = 0.6 mm] plot[ error bars/.cd, y dir=both, y explicit] coordinates {
            (5.2,0.4641) +- (0,0.0035)};  
\addplot[gray,mark=*,mark size = 0.6 mm] plot[ error bars/.cd, y dir=both, y explicit] coordinates {
            (5.8,0.4679) +- (0,0.0013)};  
\addplot[black,mark=*,mark size = 0.6 mm] plot[ error bars/.cd, y dir=both, y explicit] coordinates {
            (6.2,0.4647) +- (0,0.0035)};  
\addplot[red,mark=*,mark size = 0.6 mm] plot[ error bars/.cd, y dir=both, y explicit] coordinates {
            (7,0.4644) +- (0,0.0035)};  
\addplot[red, dashed]
    coordinates {(0.2,0.4644) (7.6,0.4644)};
\addplot[black, dashed]
    coordinates {(0.2,0.44) (7.6,0.44)};
\fi
\end{axis}
\end{tikzpicture}
\caption{The various values for $\gamma_2$ from \cref{tab:gamma_2_results}, in black, along with the mean value, in red. The direct calculations, $\gamma_2^\mathrm{d}$, not taking the experimental offset, $\gamma_0$, into account are shown in gray. The dashed black line shows the HCCM value from \cref{eq:gamma_2_model}.}
\label{fig:gamma_2_values}
\end{figure}
The resulting values are displayed in \cref{tab:gamma_2_results}, along with the MRI scanner field strength and the probe polarization configuration. We find the mean of these values to be
\begin{align}
    \gamma_2 = 0.4644(35)\:\mathrm{MHz/T^2}.
    \label{eq:gamma_2_new}
\end{align}
For the uncertainty we simply take the lowest of the uncertainties from \cref{tab:gamma_2_results}, like for \cref{eq:gamma_1_new}. Also shown in \cref{tab:gamma_2_results} are values, $\gamma_2^\mathrm{d}$, directly calculated, without taking the measurement offset, $\gamma_0$, into account. The data in \cref{tab:gamma_2_results} is shown in \cref{fig:gamma_2_values}. Since this is the first ever measurement of this number, we do not have any other experimental data to compare it to. We notice that it is on the same order of magnitude as the HCCM value found in \cref{eq:gamma_2_model}, also shown in \cref{fig:gamma_2_values}, however with a significant discrepancy of 7 standard deviations.

\section{Conclusion}
We have in this work investigated the cesium D\textsubscript{2} transition involving the extreme angular momentum states. The magnetic field dependence of this transition has been mapped with unprecedented accuracy compared to any other alkali optical transition, enabling accurate optical magnetometry at high magnetic fields.

It is very interesting to note that the excited state g-factor calculated in \cref{eq:g-factor_measured} is not in agreement with the RS coupling value of \cref{eq:g-factor_RS_free}, as we find a discrepancy of 12 ppm, significant to 31 standard deviations. According to \cite{Arimondo2016}, all measurements on alkali non-$S$ states have, until now, been in agreement with the RS value. For $D$ states in cesium discrepancies up to 55 ppm are predicted theoretically using more advanced methods \cite{Gossel2013}. No theoretical predictions are published yet for the $P$ states.

It is also very interesting to note that the quadratic diamagnetic shift is not in agreement with the HCCM value. Most of the data presented in \cite{Otto2002} show very good agreement with the HCCM value; only for potassium discrepancies of up to 2.7 standard deviations are reported. Here we report a highly significant discrepancy of 7 standard deviations.

These results could motivate theoretical work on high accuracy calculations, beyond the RS coupling scheme and the HCCM assumption.

\section{Outlook}
It should be noted that similar measurements can be made for other transitions, and other alkali atoms. This could enable high accuracy magnetometry with other laser wavelengths or other alkali atoms, and provide more data for testing of atomic structure models. In future studies, possibly involving many different field strengths, the possibility of the simple model described by \cref{eq:Freq_shift_in_practise} being insufficient should also be considered. This may become relevant for very accurate measurements, or very high field strengths. For a discussion of possible future improvements to this kind of experiment see Appendix F.

We are currently working to develop the system presented in this work, into a fully functional high speed magnetometer, by implementing continuous tracking of the magnetic frequency shift, \cref{eq:Freq_shift_in_practise}. Such a magnetometer could have applications in MRI, as described in \cite{Barmet2008}, as well as other areas where strong magnetic fields need to be stabilized or monitored. For a discussion on measurement strategies see Appendix G.

\section*{Acknowledgments}
We would like to thank Michael Zugenmaier for useful suggestions on the probe design, Axel Boisen for designing and building the heating lasers, Jan Ole Pedersen for assisting with the MRI systems, and Lars G. Hanson and Ennio Arimondo for helpful discussions.

This project has received funding from the Danish Quantum Innovation Center (QUBIZ)/Innovation Fund Denmark, the European Union’s Horizon 2020 research and innovation programme under grant agreement No 820393 (macQsimal), and Villum Fonden under a Villum Investigator Grant, grant no. 25880. The 7 T scanner was donated by the John and Birthe Meyer Foundation and The Danish Agency for Science, Technology and Innovation (grant no. 0601-01370B).

\section*{Appendix A: Probe Design}
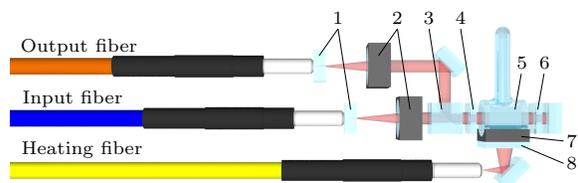
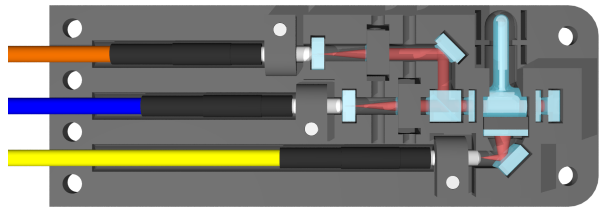
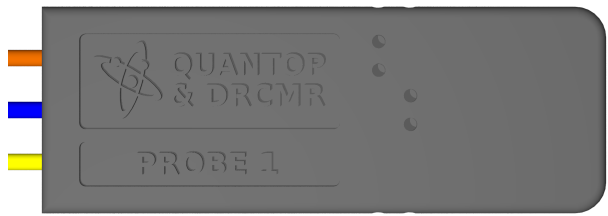
\begin{figure}[ht]
\subfloat[1: AR coated windows. 2: Lenses glued into cubic nylon holders. 3: PBS. 4: Quarter waveplate. 5: Vapor cell. 6: Optical filter with \mbox{40 \%} transmission at 852 nm. 7: Heat conducting silicone. \mbox{8: Optical} filter with \mbox{1.2 \%} transmission at 808 nm. \label{fig:Probe_b}]{
  \begin{tikzpicture}
        \node[anchor=south west,inner sep=0] (image) at (0,0) {\includegraphics[width=0.97\linewidth]{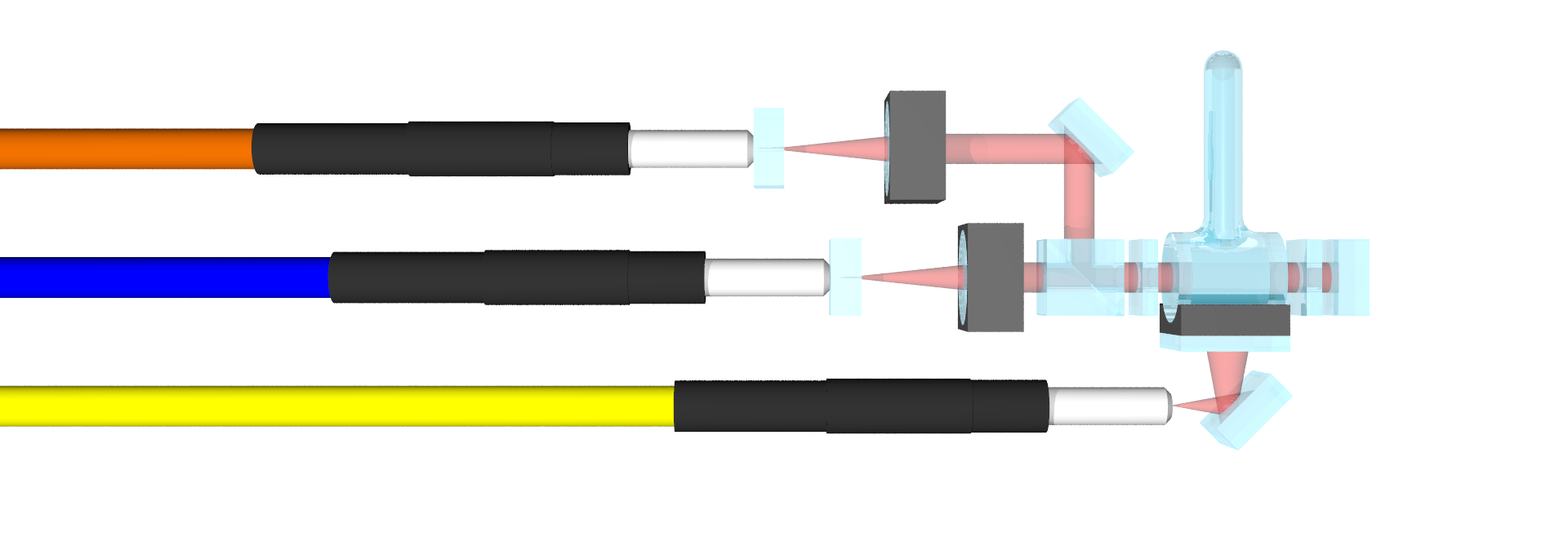}};
        \begin{scope}[x={(image.south east)},y={(image.north west)}]
        
            \node at (0.11,0.82) {\scriptsize Output fiber};
            \node at (0.1,0.59) {\scriptsize Input fiber};
            \node at (0.115,0.36) {\scriptsize Heating fiber};
            
            \node at (0.520,0.95) {\scriptsize 1};
            \draw [line width=0.05mm,-](0.520,0.90) -- (0.540,0.54);
            \draw [line width=0.05mm,-](0.520,0.90) -- (0.490,0.78);
            
            \node at (0.614,0.95) {\scriptsize 2};
            \draw [line width=0.05mm,-](0.614,0.90) -- (0.634,0.54);
            \draw [line width=0.05mm,-](0.614,0.90) -- (0.584,0.78);
            
            \node at (0.664,0.95) {\scriptsize 3};
            \draw [line width=0.05mm,-](0.664,0.90) -- (0.684,0.54);
            
            \node at (0.714,0.95) {\scriptsize 4};
            \draw [line width=0.05mm,-](0.714,0.90) -- (0.734,0.54);
            
            \node at (0.811,0.75) {\scriptsize 5};
            \draw [line width=0.05mm,-](0.810,0.70) -- (0.80,0.54);
            
            \node at (0.847,0.75) {\scriptsize 6};
            \draw [line width=0.05mm,-](0.846,0.70) -- (0.836,0.54);
            
            \node at (0.89,0.40) {\scriptsize 7};
            \draw [line width=0.05mm,-](0.877,0.40) -- (0.81,0.42);
            
            \node at (0.89,0.30) {\scriptsize 8};
            \draw [line width=0.05mm,-](0.877,0.30) -- (0.81,0.38);
            
        \end{scope}
    \end{tikzpicture}
}

\subfloat[The whole setup is put in a nylon holder. Mirrors, PBS, and windows are glued in. The fibers are clamped by the ceramic ferrules, with nylon bolts from below. Lens holders are fitted tightly into theirs slots.\label{fig:Probe_c}]{
  \includegraphics[width=0.97\linewidth]{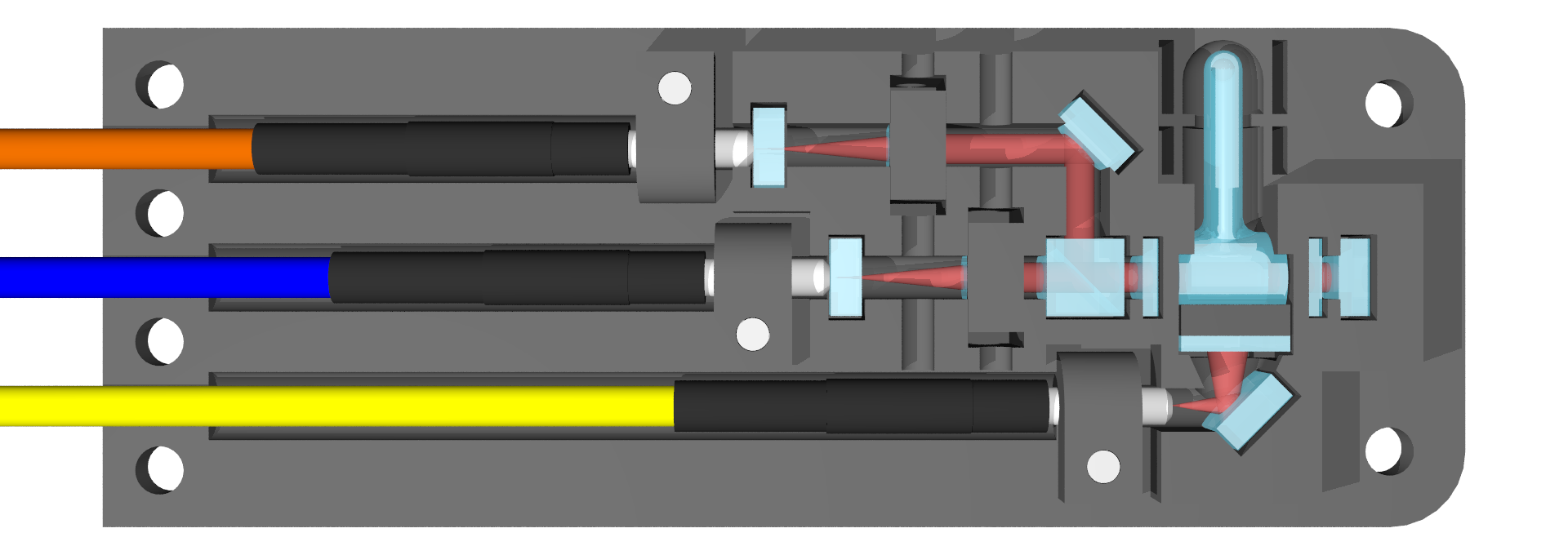}
}

\subfloat[A protective cover is fastened on top of the holder with nylon bolts from the bottom. Lens positions can be adjusted by pushing through the holes from the outside.\label{fig:Probe_d}]{
  \includegraphics[width=0.97\linewidth]{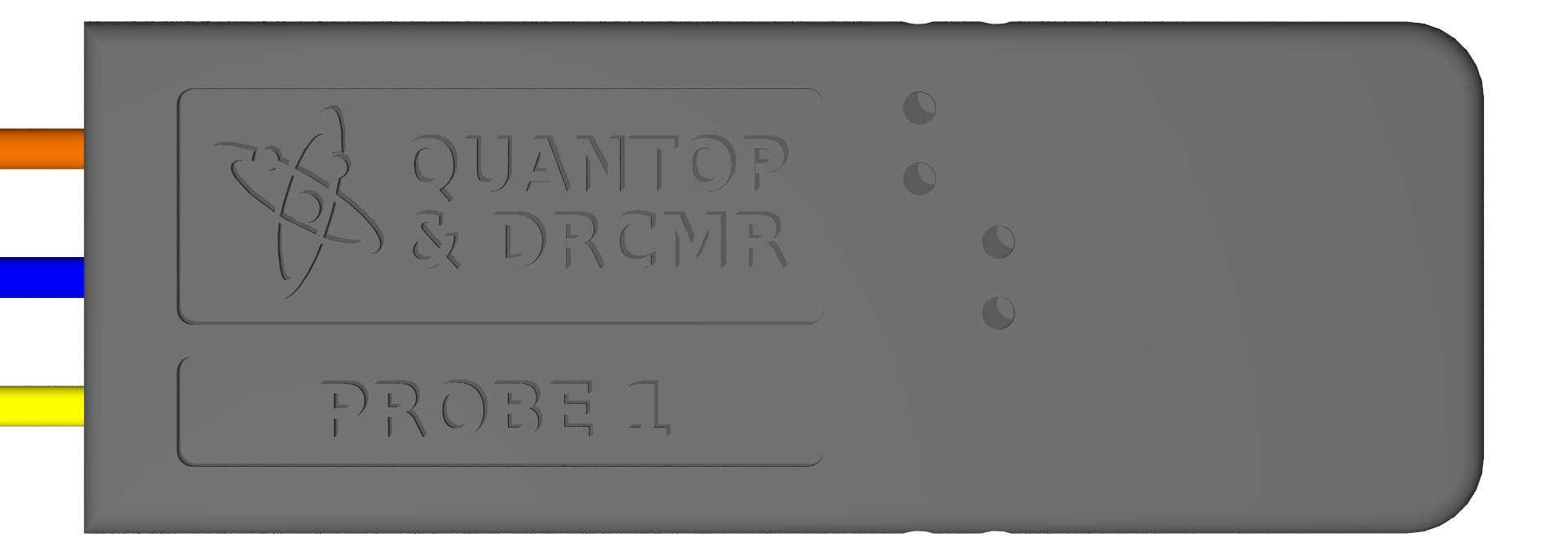}
}
\caption{The optics of the probe, and how it is mounted inside the nylon holder.}
\label{fig:Probe}
\end{figure}

The MRI-compatible fiber-coupled probe design is shown in \cref{fig:Probe}.
After the 852 nm probe light emerges from the blue PM fiber, with an angle of 4.9\textdegree, the beam is collimated by a lens, with a focal length $f = 10$ mm, to have a waist of 0.86 mm. After passing through the polarizing beam splitter (PBS), the quarter waveplate, and the vapor cell, the beam is cut by an aperture of 2 mm diameter. This means that the measurement volume of the probe is bounded by a cylinder of 2 mm diameter and 5 mm length. The return beam intensity is attenuated to \mbox{16 \%}, by passing an optical filter twice, to avoid excessive power broadening of the saturated absorption signal. Another $f = 10$ mm lens focuses the beam into a MM fiber with a core diameter of 0.4 mm. The 808 nm heating laser light is delivered through a similar MM fiber. The heating laser beam is absorbed in an optical filter and a silicone heat conductor transfers the heat to the vapor cell, allowing for very localized heating. With the stem of the vapor cell pointing away from the point of heating, it is ensured that the coldest point of the cell is far away from the probing beam path, such that cesium does not condense on the windows and block the probe beam. The two probe beam input and output fibers are terminated by windows with anti-reflection (AR) coating on the side facing away from the fibers, to reduce spurious etalon fringes from the fibers in the spectrum. Index matching gel is applied at the interface between the fiber tips and the windows. The fibers are 19 m long. Mirrors, windows and PBS are fastened with glue. The lenses are glued into cubic holders which are mounted by a tight fit in their slots. 

The nylon enclosure measures $90\times 33\times 10\: \mathrm{mm}^3$, and is 3D printed using HP Multi Jet Fusion (MJF) technology. While stereolithographic (SLA) 3D printing produces very nice results, as demonstrated in \cite{Madkhaly2021}, the material is not compatible with temperatures approaching 100 \textdegree C, like nylon, which is why the MJF method is used here.

By pushing through holes from the outside the lens positions can be adjusted to optimize beam direction and output fiber coupling. We achieve a fiber coupling close to \mbox{100 \%}, reasonably stable during daily handling.

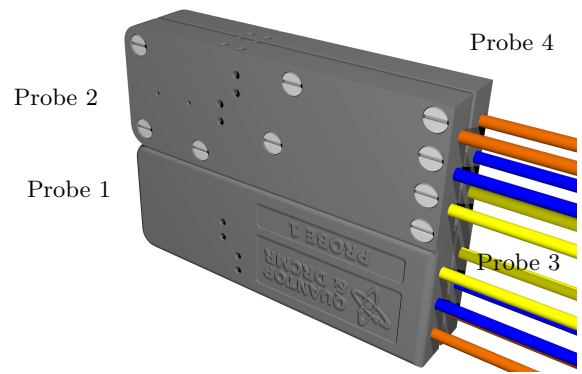
\begin{figure}[htb]
    \centering
    \begin{tikzpicture}
        \node[anchor=south west,inner sep=0] (image) at (0,0) {\includegraphics[width=\linewidth]{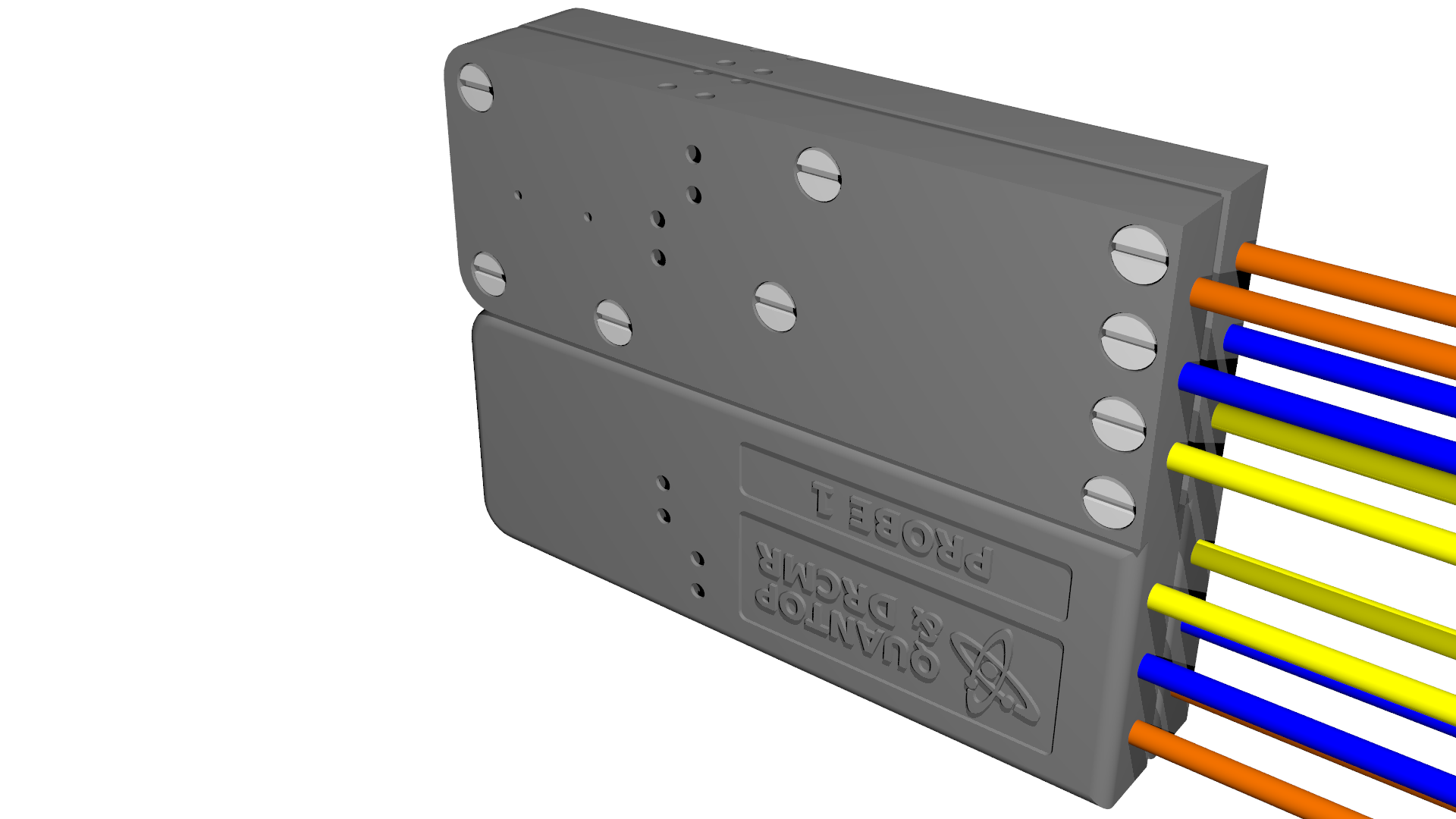}};
        \begin{scope}[x={(image.south east)},y={(image.north west)}]
        \node at (0.22,0.50) {Probe 1};
        \node at (0.20,0.75) {Probe 2};
        \node at (0.91,0.30) {Probe 3};
        \node at (0.90,0.90) {Probe 4};
        \end{scope}
    \end{tikzpicture}
    \caption{The four probes in the configuration used inside the scanner. They are held together by two thin sewing threads (not shown in the image).}
    \label{fig:Probes Configuration}
\end{figure}

During the experiments described in this work the four probes are strapped together, by two thin sewing threads, in a configuration with probe 1 and 3 below, and probe 2 and 4 on top, as shown in \cref{fig:Probes Configuration}.
When the spherical water container is removed from the MRI scanner during a data acquisition run, the probes are positioned, as well as possible, at the center of where the sphere used to be, suspended from a thin sewing thread. The sewing threads are assumed to have negligible effect on the magnetic field.

Probe 1 and 3 are heated with 600 mW of optical power, and probe 2 and 4 are heated with 500 mW. The reference probe is heated with 250 mW. After about an hour the probe temperatures stabilize around 43~\textdegree C, and the reference temperature stabilize around 35~\textdegree C. This corresponds to a density of about $26\times 10^{16}\:\mathrm{m}^{-3}$ for the probes and $13\times 10^{16}\:\mathrm{m}^{-3}$ for the reference. Compared to about $3\times 10^{16}\:\mathrm{m}^{-3}$ at room temperature, this gives a significant increase in absorption depth, and hence signal strength.

About 542 \textmu W of optical probe power is sent to the probes, and about 175 \textmu W is sent to the reference. Notice that for the probes only a smaller fraction of the probe light is actually resonant, as the sidebands generated by the EOM are the ones used. At 3 T 3\textsuperscript{rd} sidebands are used, and at 7 T 5\textsuperscript{th} sidebands are used. This means that only about 23 \% and 14 \% of the probe light is on resonance, respectively.

\section*{Appendix B: Magnetic Field Shift of the Probes}
The shielded proton gyromagnetic ratio is defined such that the field that is measured is the one in the vacuum left when the spherical container is removed. When the sphere is removed, and replaced with the four probes, and air, the magnetic field is changed slightly due to the change in magnetic susceptibility. This is taken into account by the factor $\sigma$ introduced in \cref{eq:Freq_shift_in_practise}. We have measured the volume magnetic susceptibility, $\chi$, for all the components of the probes, by a method similar to that described in \cite{Wapler2014}: A sample of the component material is submerged in water, and the field distortion around the sample is mapped using the 7 T MRI scanner. From a fit the difference in sample and water magnetic susceptibility, $\chi-\chi_\mathrm{H_2O}$, is determined. The sample magnetic susceptibility is then found using $\chi_\mathrm{H_2O} = -9.032$ ppm \cite{Wapler2014}. For components easily isolated the results are shown in \cref{tab:Susceptibilities_materials}.
\begin{table*}[htb]
\centering
\subfloat[Measurements on parts consisting of only a single material. The method is estimated to be accurate to about 5 \% for the value of $\chi-\chi_\mathrm{H_2O}$.]{
    \begin{tabular}{l l r r}
    \hline\hline
    Material & Component & $\chi-\chi_\mathrm{H_2O}$ (ppm) & $\chi$ (ppm) \\
    \hline
    Unknown & Heating filter & 18.5 & 9.4 \\
    MJF Nylon PA12 & 3D print & 0.121 & -8.911 \\
    Nylon 66 UL94V-2 & Bolts & -0.213 & -9.245 \\
    Synthetic quartz & Waveplate & -4.34 & -13.37 \\
    Schott Borofloat 33 & Mirrors and windows & -2.13 & -11.17\\
    Schott N-BK7 & Lenses & -3.31 & -12.35 \\
    Schott N-SF1 & PBS & -0.524 & -9.556 \\
    Schott NG4 & Probe beam filter & 149 & 140 \\
    Silicone, RS 174-5694 & Heat conductor & -1.27 & -10.31\\
    Zirconia & Fiber ferrules & -1.05 & -10.09\\
    \hline\hline
\end{tabular}
\label{tab:Susceptibilities_materials}}
\quad
\subfloat[Measurements on compound parts not easily separated in single materials. The method is estimated to be accurate to about 25 \% for the value of $\chi-\chi_\mathrm{H_2O}$.]{
\begin{tabular}{l r r} \\ \\ \\ \\
\hline\hline
Component & $\chi-\chi_\mathrm{H_2O}$ (ppm) & $\chi$ (ppm) \\
\hline
Fiber cable (heat) & 1.38 &  -7.65 \\
Fiber cable (in) & 2.45 &  -6.58 \\
Fiber cable (out) & 1.12 &  -7.91 \\
Fitting (heat) & 4.61 & -4.42 \\
Fitting (in) & 2.22 &  -6.82 \\
Fitting (out) & 4.58 &  -4.45 \\
\hline\hline
\end{tabular}
\label{tab:Susceptibilities_compounds}}
\caption{Measured volume magnetic susceptibilities.}
\end{table*}
We note that the values for $\chi-\chi_\mathrm{H_2O}$ found for Schott Borofloat 33 and Schott N-BK7 agrees with the values from \cite{Wapler2014} within about 5 \%. Hence this is chosen as the level of uncertainty. This implies that magnetic susceptibilities close to $\chi_\mathrm{H_2O}$ are measured much more accurately than those far away. Notice how the optical filters are highly paramagnetic; especially the probe filter.

For the fiber cables (i.e.\ the outer jacket of the cable and everything inside it, including air) and the fittings (which are keeping the fiber cables and the fiber ferrules together), ``effective'' magnetic susceptibilities are determined, for the compound components. The results are shown in \cref{tab:Susceptibilities_compounds}. Since this is measured without disassembling the fiber-cables, the method is less accurate. Detailed knowledge of these components is fortunately not very important for the field shift at the position of the vapor cell.

For the vapor cell, a model is constructed using the value for quartz glass, from \cite{Wapler2014}, ($\chi = -11.27$ ppm) and vacuum ($\chi = 0$). The field shift of this model is seen to match reasonably with the measured field shift from a vapor cell. Our vapor cells are made from Momentive GE214 Fused Quartz (body), and Corning 7980 Fused Silica (windows), by Precision Glass Blowing.

The magnetic susceptibility of the surrounding air is taken to be 0.36 ppm, as in \cite{Schenck1996}.

The magnetic susceptibility of cesium is calculated from the data in \cite{Haynes2016} to be about 5.1 ppm. Hence the very small amounts of cesium sitting in the bottom of the vapor cell stem can safely be ignored. Similarly the glue used to hold the optical elements in place, the index matching gel, and the thin optical coatings are assumed to be negligible.

\begin{figure}[htb]
    \centering
    \includegraphics[width=0.9\linewidth]{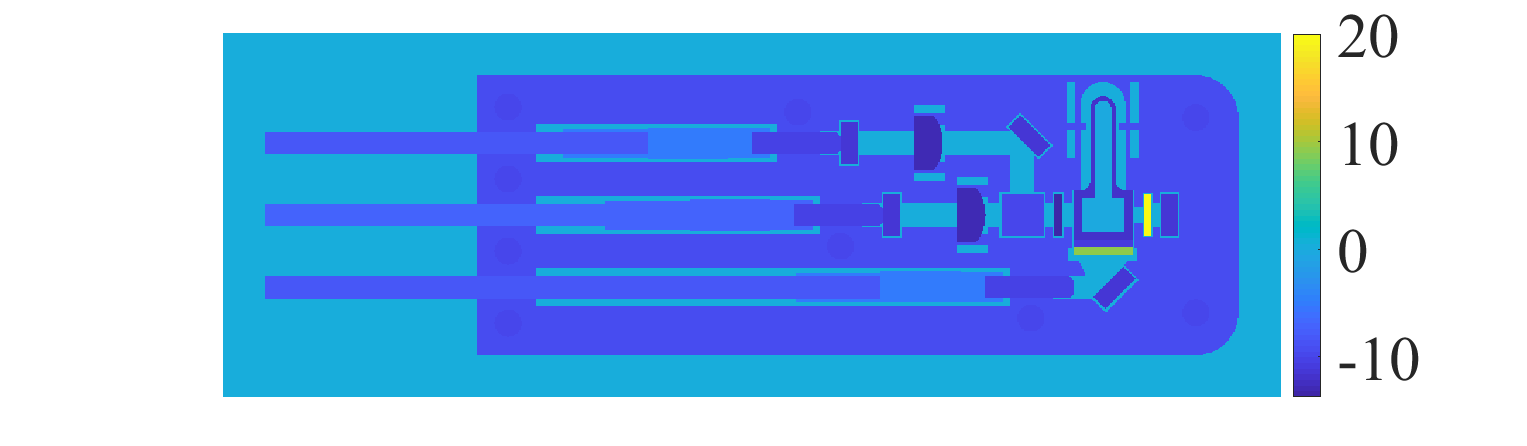}
    \caption{A cross section of the 3D susceptibility model of the probe. The cross section is made through the center of the vapor cell. We cap the color scale at 20 ppm, even though the probe filter susceptibility is 140 ppm, to highlight the details of the entire structure.}
    \label{fig:Susceptibility_map}
\end{figure}

\begin{figure}[htb]
    \centering
    \includegraphics[width=0.9\linewidth]{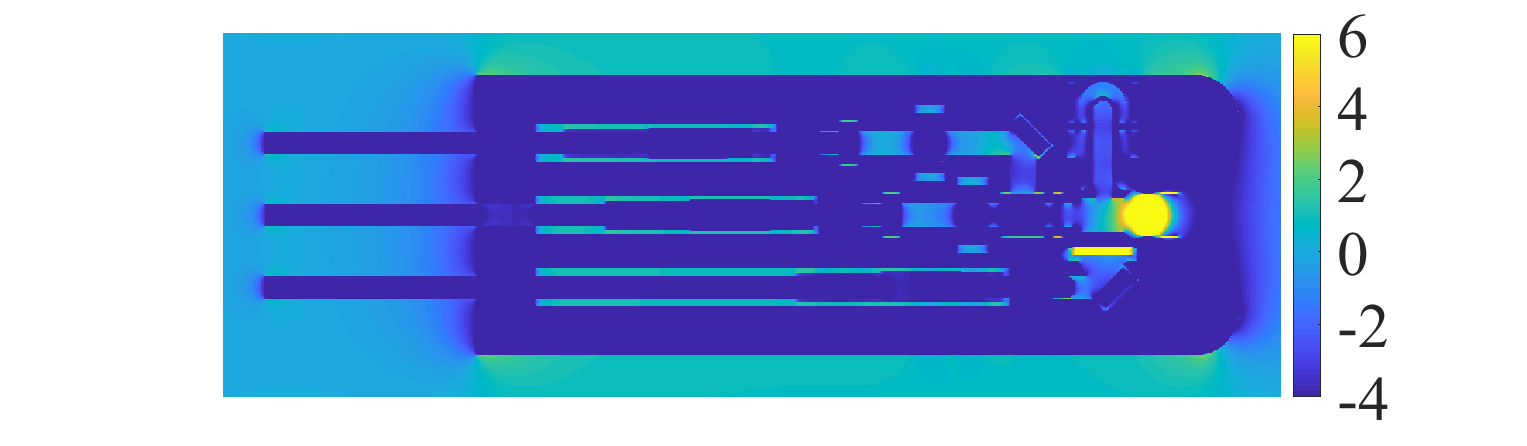}
    \caption{Calculated field shift map. The cross section is the same as in \cref{fig:Susceptibility_map}. The effect of the rest of the fibers, not included in this model, is verified, through a similar simulation, to have a negligible effect, at the position of the vapor cell. We cap the color scale to match the distribution inside the vapor cell.}
    \label{fig:Field_shift_map}
\end{figure}

Using all the above, a 3D susceptibility model of the probe is constructed, as seen in \cref{fig:Susceptibility_map}.
Using the method described in \cite{Bouwman2012,Bouwman2012_script}, we then calculate the field shift caused by this distribution of magnetic susceptibility. This method takes into account the effect of the Lorentz sphere, i.e.\ the effect of the granular structure of matter \cite{Koch2006,Haacke2014}. Since we care about the field shift in the vacuum inside the vapor cell, we find the ``continuous matter-field shift'' by adding $\frac{2}{3}\chi$ to the calculated field shift. Since the calculation is done on the susceptibility relative to the surrounding air, we also add $\frac{2}{3}\chi_\mathrm{Air}$ to account for the sphere of air that replaces the sphere of water. The resulting field shift map is shown in \cref{fig:Field_shift_map}.
It should be noted that including the Lorentz sphere actually also works, since in this case you do not have any field shift from the sphere of air, that surrounds the probe.

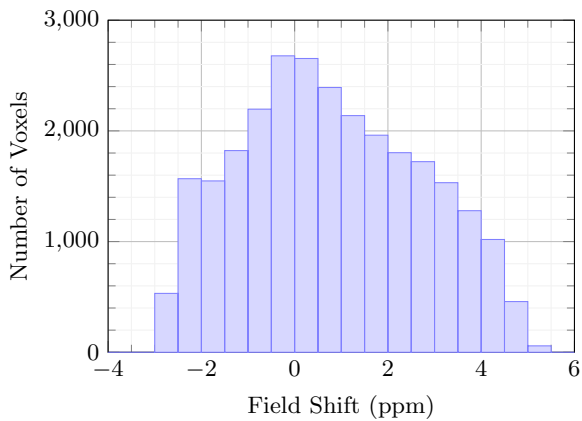
\begin{figure}[htb]
\centering
\begin{tikzpicture}
\begin{axis}[
    height              = 6 cm,
	width               = 0.9\linewidth,
	xlabel              = Field Shift (ppm),
	ylabel              = Number of Voxels,
	xmin                = -4,
	xmax                = 6,
	ymin                = 0,
    ymax                = 3000,
	grid                = both,
	grid style          = {line width=.4pt, draw=gray!10},
    major grid style    = {line width=.4pt, draw=gray!50},
	minor x tick num    = 3,
	minor y tick num    = 4,
	clip                = true
    ]
\ifNotDraft
\addplot [ybar interval,mark=no,color=my_light_blue,fill=my_lighter_blue] table[y=N,x=Edges] {Histogram_beam_path.txt};
\fi
\end{axis}
\end{tikzpicture}
\captionsetup{width=10 cm}
\caption{Distribution of voxels in the beam path inside the vapor cell, according to the simulated field shift. Notice how the highly paramagnetic optical filter next to the vapor cell creates a magnetic field gradient along the beam path.}
\label{fig:Histogram_beam_path}
\end{figure}

Picking out the voxels that make up the probe beam path inside the vapor cell, we find a distribution of field shifts as seen in \cref{fig:Histogram_beam_path}.
Varying the voxel size, the surrounding zero-padding, the magnetic susceptibilities of components close to the vapor cell, and the exact position of the highly paramagnetic probe filter, we find the mean value for the field shift in the beam path inside the vapor cell to be $1 + 0.49(50)\times 10^{-6}$. The uncertainty on this number also takes into account the variation over the radius of the beam, since it is not clear if the center or the edge of the beam contributes the most to the measured signal. The main contribution to the uncertainty is related to the uncertainty on the probe filter susceptibility and its exact position. A similar simulation is performed on all four probes strapped together as seen in \cref{fig:Probes Configuration}, to give 
\begin{align}
    \sigma = 1 + 0.93(50)\times 10^{-6}.
\end{align}

\section*{Appendix C: Magnetic Field Determination by proton NMR Spectroscopy}
Accurate magnetic field determinations in the tesla range, uses that the \textit{shielded proton gyromagnetic ratio} is known with very high accuracy \cite{Phillips1977,Mohr2000,Tiesinga2021}, as
\begin{align}
    \gamma'_\mathrm{p}(25\text{ \textdegree C}) = 42.576\:384\:74(46) \:\mathrm{MHz/T}.
\end{align}
This value refers to hydrogen nuclei (protons) in a spherical container of pure water, at 25 \textdegree C. The water shielding factor depends slightly on the temperature, $t$, \cite{Petley1984,Mohr2000}, as
\begin{align}
    \frac{\gamma'_\mathrm{p}(t)}{\gamma'_\mathrm{p}(25\text{ \textdegree C})} = 1-10.36(30)\times 10^{-9}\text{ \textdegree C}^{-1}(t-25\text{ \textdegree C}).
\end{align}
By nuclear RF excitation, and subsequent readout of the free induction decay (FID), the proton precession frequency, $\nu_\mathrm{p}$, is measured. The magnetic field can be then be found as
\begin{align}
    B = \frac{\nu_\mathrm{p}}{\gamma'_\mathrm{p}(t)}.
\end{align}
The magnetic field, $B$, is defined as the field in the vacuum left when the sphere of water is removed. Notice that the material of the container is not important, as the field shift inside a spherical shell is zero \cite{Foot2012}. 

For our case, where we use an MRI scanner, the acquired RF signal is first mixed down by a frequency, $\nu_0$, chosen by the scanner, close to the actual resonance frequency. The measured frequency of the FID is then $\nu_\mathrm{m}$, such that $\nu_\mathrm{p} = \nu_0 + \nu_\mathrm{m}$. We find $\nu_\mathrm{m}$ as the peak value of the Fourier transform of the downmixed FID signal. The uncertainty is estimated by inspection of a field image, of the spherical water container, produced by the scanner. For the 3 T scanner the full width at 25 \% of the peak value is found to be a good estimate for the uncertainty. For the 7 T scanner the full width at 5 \% of the peak value is used. For the six data acquisition runs we note $\nu_0$ and $\nu_\mathrm{m}$ in \cref{tab:NMR_spectroscopy_data}.
\begin{table}[ht]
    \centering
    \begin{tabular}{l l r r r}
    \hline\hline
    $B$ & Config. & $\nu_0$ (Hz) & $\nu_{\mathrm{m},a}$ (Hz) & $\nu_{\mathrm{m},b}$ (Hz)  \\
    \hline
    3 T & $\scriptstyle --++$   & 127\:778\:089 & 4(36)  & 13(38)   \\
    3 T & $\scriptstyle ++--$   & 127\:777\:868 & 5(32)  & 34(41)   \\
    7 T & $\scriptstyle ++--$   & 298\:037\:729 & 3(60)   & 5(61)   \\
    7 T & $\scriptstyle --++$  & 298\:037\:737 & -13(42) & -14(65)  \\
    7 T & $\scriptstyle --++^*$ & 298\:037\:744 & -12(74) & -12(72) \\
    7 T & $\scriptstyle ++--^*$ & 298\:037\:724 & -8(62)  & -4(61)  \\
    \hline\hline
    \end{tabular}
    \caption{Downmixing, $\nu_0$, and peak frequencies, $\nu_\mathrm{m}$, for the six data acquisition runs.}
    \label{tab:NMR_spectroscopy_data}
\end{table}
From this the proton precession frequencies in \cref{tab:final_data} follows. Notice that imperfections in the spherical shape, such as container deformations, residual air bubbles, and the small hole used for water filling, are taken into account by this uncertainty estimation. The sphere has a diameter of 100 mm, and hence covers a much larger volume than the four vapor cells.

\section*{Appendix D: Measuring optical frequency shifts}
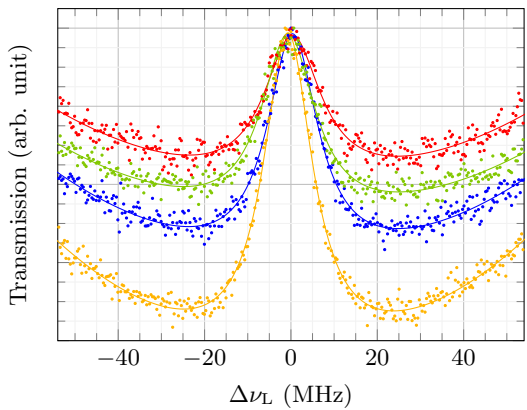
\begin{figure}[thb]
  \centering
  \begin{tikzpicture}
  \begin{axis}[
			height              = 6 cm,
			width               = 0.9\linewidth,
			xlabel              = $\Delta\nu_\mathrm{L}$ (MHz),
			ylabel              = Transmission (arb. unit),
			scaled y ticks      = false, 
			yticklabels         = {},
			xmin                = -54,
			xmax                = 54,
			ymin                = -0.008,
			ymax                = 0.0005,
			grid                = both,
			grid style          = {line width=.4pt, draw=gray!10},
            major grid style    = {line width=.4pt, draw=gray!50},
			minor x tick num    = 3,
			minor y tick num    = 3,
			clip                = true]
  \ifNotDraft
  \addplot [color=blue,   only marks,mark size=0.4] table[x=Freq,y=avgA]{Lorentzian_fits.txt};
  \addplot [color=red,    only marks,mark size=0.4] table[x=Freq,y=avgB]{Lorentzian_fits.txt};
  \addplot [color=my_green,  only marks,mark size=0.4] table[x=Freq,y=avgC]{Lorentzian_fits.txt};
  \addplot [color=my_yellow, only marks,mark size=0.4] table[x=Freq,y=avgD]{Lorentzian_fits.txt};
  \addplot [color=blue]   table[x=Freq,y=f_A(freq)]{Lorentzian_fits.txt};
  \addplot [color=red]    table[x=Freq,y=f_B(freq)]{Lorentzian_fits.txt};
  \addplot [color=my_green]  table[x=Freq,y=f_C(freq)]{Lorentzian_fits.txt};
  \addplot [color=my_yellow] table[x=Freq,y=f_D(freq)]{Lorentzian_fits.txt};
  \fi
  \end{axis}
  \end{tikzpicture}
  \caption{Average of 100 laser frequency scans across the point in between the $+7$ T and the $-7$ T transition. The 5\textsuperscript{th} upper sideband probes the $+7$ T transition, in probe 1 and 2 (blue and red data); and the 5\textsuperscript{th} lower sideband probes the $-7$ T transition, in probe 3 and 4 (green and yellow data, respectively).}
  \label{fig:Lorentzian_fits}
\end{figure}

\begin{figure}[htb]
  \centering
  \begin{tikzpicture}
  \begin{axis}[
			height              = 6 cm,
			width               = 0.9\linewidth,
			xlabel              = $10\cdot\Delta\nu_\mathrm{EOM}$ (MHz),
			ylabel              = $\nu_i-\nu_1$ (MHz),
			scaled x ticks      = false,
			xticklabel style    ={  /pgf/number format/fixed,
                                    /pgf/number format/precision=1,
                                    /pgf/number format/fixed zerofill,
                                    /pgf/number format/1000 sep={}},
			xmin                = 195921.4,
			xmax                = 195923.4,
			ymin                = -1.1,
			ymax                = 1.1,
			grid                = both,
			grid style          = {line width=.4pt, draw=gray!10},
            major grid style    = {line width=.4pt, draw=gray!50},
			minor x tick num    = 4,
			minor y tick num    = 2,
			clip                = true]
  \ifNotDraft
  \addplot [color=blue,   only marks,mark size=1] table[x=Freq,y=nu_A]{Linear_fits.txt};
  \addplot [color=red,    only marks,mark size=1] table[x=Freq,y=nu_B]{Linear_fits.txt};
  \addplot [color=my_green,  only marks,mark size=1] table[x=Freq,y=nu_C]{Linear_fits.txt};
  \addplot [color=my_yellow, only marks,mark size=1] table[x=Freq,y=nu_D]{Linear_fits.txt};
  
  \addplot [color=blue]   table[x=Freq,y=linear_A]{Linear_fits.txt};
  \addplot [color=red]    table[x=Freq,y=linear_B]{Linear_fits.txt};
  \addplot [color=my_green]  table[x=Freq,y=linear_C]{Linear_fits.txt};
  \addplot [color=my_yellow] table[x=Freq,y=linear_D]{Linear_fits.txt};
  
  \addplot[color=blue,mark size=1] plot[ error bars/.cd, y dir=both, y explicit] coordinates {
            (195921.8,0) +- (0,0.0727)};
  \addplot[color=red,mark size=1] plot[ error bars/.cd, y dir=both, y explicit] coordinates {
            (195921.8,-0.1782) +- (0,0.1191)};
  \addplot[color=my_green,mark size=1] plot[ error bars/.cd, y dir=both, y explicit] coordinates {
            (195921.8,-0.4731) +- (0,0.0749)};
  \addplot[color=my_yellow,mark size=1] plot[ error bars/.cd, y dir=both, y explicit] coordinates {
            (195921.8,-0.6356) +- (0,0.0487)};
  \fi
  \end{axis}
  \end{tikzpicture}
  \caption{Line center $\nu_i$, for probes $i=\{1,2,3,4\}$, relative to $\nu_1$, as a function of $10\cdot\nu_\mathrm{EOM}$. Colors are as in \cref{fig:Lorentzian_fits}. The leftmost data points correspond to the line centers found from the fits in \cref{fig:Lorentzian_fits}. Error bars are found as 68 \% confidence intervals, i.e.\ one standard deviation, as detailed in \cite{CurveFittingToolboxUserGuide}. To keep the figure clear and readable, error bars are shown only for a single representative line of data.}
  \label{fig:Linear_fits}
\end{figure}
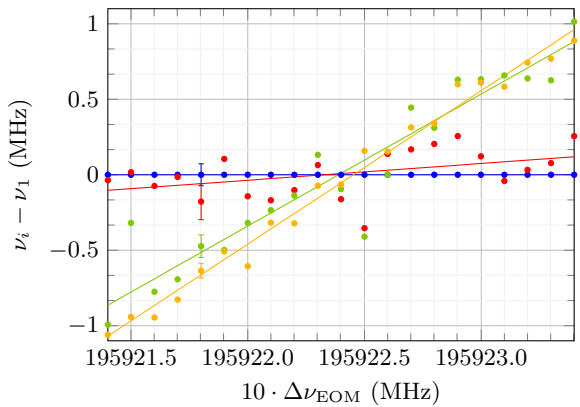

Central to this work is the method to accurately measure a resonance frequency difference by sideband spectroscopy. By choosing an EOM modulation frequency, $\nu_\mathrm{EOM}$, equal to an integer fraction of the frequency difference the saturated absorption resonances, as probed by the carrier or sidebands, can be brought to overlap. To do this in a systematic and unbiased way, a series of different frequencies, $\nu_\mathrm{EOM}$, are tried. In steps of 0.01 MHz, a range of 0.20 MHz is covered. 100 laser frequency scans are averaged for each $\nu_\mathrm{EOM}$. As an example we take the calculation of the shift $\Delta\nu_{\pm B}$ in the last line in \cref{tab:final_data}, i.e.\ in the 7 T scanner, with probe 1 and 2 configured with $\sigma_+$ polarization, probe 3 and 4 with $\sigma_-$ polarization. In \cref{fig:Lorentzian_fits} is shown a measurement (100 averages) with $\nu_\mathrm{EOM} = 19\:592.14$ MHz. 
A fit of a second degree polynomial background and a Lorentzian line shape, is performed for each probe. The frequency axis is estimated in a prior scan across the 0 T spectrum from the reference probe. The exact scaling of this is not important. Changing $\nu_\mathrm{EOM}$ in steps of 0.01 MHz up to 19592.34 MHz, we get a series of fitted line-centers as shown in \cref{fig:Linear_fits}.
Straight lines are fitted to the data, and the four intersections are found. The optical frequency shift is then found as the average of these to be $\Delta\nu_{\pm B} = 195\:922.427$ MHz. Similarly $\Delta\nu_{+B}$ and $-\Delta\nu_{-B}$ are found by sweeping $\nu_\mathrm{EOM}$ across the point where the $+7$ T and $-7$ T resonances overlap with the 0 T resonance, respectively. Here only two intersections are found, since there is only one reference probe.

To estimate the uncertainty on the overlapping procedure, we note that the first measurement, $\Delta\nu_{\pm B}$, should equal the sum of the two following measurements, $\Delta\nu_{+B}$ and $-\Delta\nu_{-B}$, so the difference between those two numbers represents an uncertainty on the overlapping procedure. We find the root mean square of the differences to be 0.034 MHz.

Apart from the overlapping procedure, there is also an uncertainty on how well the fitted line shape center represents the actual resonance frequency. In particular, a significant ``geometrical shift'' has been observed, clearly correlating with the geometrical broadening (see e.g.\ \cite{Fedorov2015}) associated with the lenses not being well positioned, such that the beam is not reflected exactly 180\textdegree\ backwards through the vapor cell. All lenses are adjusted as well as possible to reduce the geometrical broadening and shift. Configuring all the probes with $\sigma_+$ polarization, we see a variation of up to 0.054 MHz between the probes. Any shift common to the probes 1-4, but differing from the reference probe is described by the constant $\gamma_0$ in \cref{eq:Freq_shift_in_practise}.

We add the two above described error sources to get 0.088 MHz. This is the uncertainty used for the optical frequency shifts in \cref{tab:final_data}. Other considered error sources are discussed in Appendix E.

\section*{Appendix E: Optical Resonance Shifting Error Sources}
A number of error sources potentially contributing to a systematic shift of the optical resonances, has been considered in this work. Most notably is the ``geometrical shift'' associated with lens positions, and the related geometrical broadening of the line. We have not observed a similar shift at zero field, making this a good candidate for explaining the measurement offset, $\gamma_0$. With poor lens adjustment, linewidths up to 30 MHz, and 0.4 MHz line center shifting has been observed. By proper lens adjustment linewidths of about 16 MHz are achieved. Since power broadening alone cannot account for such a linewidth, a residual geometrical broadening, common among the probes is assumed. Hence a related geometrical shift of about $\gamma_0 = 0.159\:\mathrm{MHz}$ is very reasonable. Also an inability of the fit (a second degree polynomial background and a Lorentzian lineshape) to nicely find the line center in the asymmetric zero field spectrum, might also contribute to $\gamma_0$. However with an estimated uncertainty of 100 \% in $\gamma_0$, these error sources should certainty be accounted for in our final result for $\gamma_2$. It is comforting that the geometrical shift seems to be the same at positive and negative field strengths, and hence the probe configuration alternations ($\scriptstyle --++ \:\leftrightarrow\: ++--$) actually removes this error source in the $\gamma_1$ result, and the derived Landé g-factor result. Unfortunately the handling of the probes in between the different measurements, might shift the lens positions slightly, so we cannot claim a complete immunity to this shift, which is why we include this error source in the uncertainty on the optical frequency shifts.

Since the probe and the reference cells are kept at different temperatures, one might also suspect that a pressure shift could contribute to $\gamma_0$. However, the temperature dependence of the line center has been investigated, experimentally, and found to be insignificant. Pressure shifts with buffer gasses are typically up to about 10 MHz/torr \cite{Pitz2010}. Since the pressure in our cells are about $10^{-5}$ torr (i.e.\ the cesium vapor pressure), we should expect pressure shifts of only up to 0.0001 MHz, if the results from buffer gasses can be applied to pure cesium.

Another significant error source that has been considered is the unavoidable higher order sidebands probing the more magnetic field sensitive transitions with $m_J = -\frac{1}{2}\leftrightarrow\frac{1}{2}$. In particular at 7 T the 8\textsuperscript{th} lower sideband, in the $\sigma_-$ configured probes, produces a weak peak that can in fact impact the line center determination. In experiments this manifests as a clear discrepancy between the first measurement, $\Delta\nu_{\pm B}$, and the sum of the two following measurements, $\Delta\nu_{+B}$ and $-\Delta\nu_{-B}$. For our measurements at 7 T we have therefore turned down the EOM drive power a bit, from where the optical power in the 5\textsuperscript{th} sideband is optimized. This way the 8\textsuperscript{th} sideband can be reduced heavily while only reducing the 5\textsuperscript{th} slightly. This error source can still not be completely removed this way, and is hence accounted for through the uncertainty in the peak overlapping procedure.

The light shift (AC Stark effect) from the sidebands not on resonance, but detuned $\delta = \nu_\mathrm{EOM}$, will shift the resonant transition slightly. From \cite{Foot2012} we have that this light shift is
\begin{align}
    \Delta\nu_\mathrm{light} = \frac{\Gamma^2 I/I_\mathrm{sat}}{4\cdot (2\pi)^2\cdot\delta},
\end{align}
where $\Gamma=\tau^{-1}$ is the decay rate of the excited state \cite{Steck2019,Young1994}. This effect will be strongest for the experiments at 3 T using the 3\textsuperscript{rd} sidebands. In this case, in the center of the beam, we have approximately $I/I_\mathrm{sat} = 10$, for the sideband \textit{on} resonance. Considering that the two neighboring sidebands are approximately half the intensity, and using $\delta= 14$ GHz, this gives a light shift of about 0.002 MHz. Since the two nearest sidebands in fact shift in opposite directions, we find this effect to be negligible. We have also, experimentally, investigated the probe power dependence of the line center, and found it to be insignificant.

As shown in \cite{Grimm1989}, the pressure of the probe light exerts a force on the atoms, that breaks the symmetry of the velocity distribution. We have not considered this effect in detail, but note that it might contribute to $\gamma_0$. In fact this may actually be related to what we here call a ``geometrical shift''.

\section*{Appendix F: Future High Accuracy Measurements}
To improve on the experiments presented in this work a number of steps can be taken:
\begin{itemize}
    \item Increasing the field strength will be useful, in particular for measuring the quadratic diamagnetic shift. MRI systems, NMR spectrometers, or custom made systems may be employed. As shown here, it can be useful to include more than one field strength in a study, when measuring the quadratic shift, whereas the linear shift can reliably be measured using only a single field strength.
    \item Improving on the field homogeneity (shimming) will be necessary to gain the most from using higher field strengths.
    \item Using higher EOM drive frequencies, such that lower order sidebands are used, could provide better a signal-to-noise ratio, and remove the problem of higher order sidebands probing higher lying transitions.
    \item Developing a more robust probe design, which can sustain higher cell temperatures, has fixed and well overlapping beams, and does not include highly paramagnetic components, which complicates magnetic field shift calculation, could also make future measurements more accurate, and possibly reduce $\gamma_0$.
    \item Alternatively, one could employ a spherical vapor cell that simply replaces the spherical water container in the setup, similar to the approach taken in the measurement of the shielded proton gyromagnetic ratio \cite{Phillips1977}. This would remove the error source introduced by the probe field shift.
\end{itemize}

\section*{Appendix G: High Field Optical Magnetometry}
We here consider two measurement methods to implement high field optical magnetometry.

The first method involves measuring the frequency shift from the resonance at 0 T to the resonance in the field, $B$. This uses a single reference probe in a magnetic shield, and any number of probes in the field, depending on the spatial resolution needed for the application. The laser frequency can be stabilized using the reference, while EOM generated sidebands are used to track the resonances from the probes in the field. Assuming that $\sigma_+$ configured probes are used, the magnetic field is calculated from the frequency shift, using \cref{eq:Freq_shift_in_practise}, as
\begin{align}
    B = \frac{-\gamma_1 + \sqrt{\gamma_1^2-4\gamma_2(\gamma_0-\Delta\nu)}}{2\gamma_2\sigma}.
\end{align}
Using the values for $\sigma$, $\gamma_0$, $\gamma_1$, and $\gamma_2$ from this work and the last optical frequency measurement, $\Delta\nu_{+B}$, from \cref{tab:final_data}, we find
\begin{align}
    B = 7.000075(18)\:\mathrm{T}.
\end{align}
That is, a measurement of the magnetic field with 2.6 ppm accuracy.

The second method involves measuring the frequency shift from $-B$ to $+B$, that is, $\Delta\nu_{\pm B}$. This method has the advantage that it is not sensitive to the diamagnetic shift and the reference offset, and that it measures about twice the frequency shift compared to first method, i.e.\ half the relative uncertainty. For these reasons it is more accurate. The disadvantage is that it requires two probes in the magnetic field, and as such only works for highly homogeneous fields. The magnetic field is calculated, using \cref{eq:linear_measurement}, as
\begin{align}
    B = \frac{\Delta\nu_{\pm B}}{2\gamma_1\sigma}.
\end{align}
Using the values for $\sigma$ and $\gamma_1$ from this work and the last optical frequency measurement, $\Delta\nu_{\pm B}$, from \cref{tab:final_data}, we find
\begin{align}
    B = 7.0000726(46)\:\mathrm{T}.
\end{align}
That is, a measurement of the magnetic field with 0.7 ppm accuracy.

\section*{Appendix H: Experimental Observation of the Line Splitting}
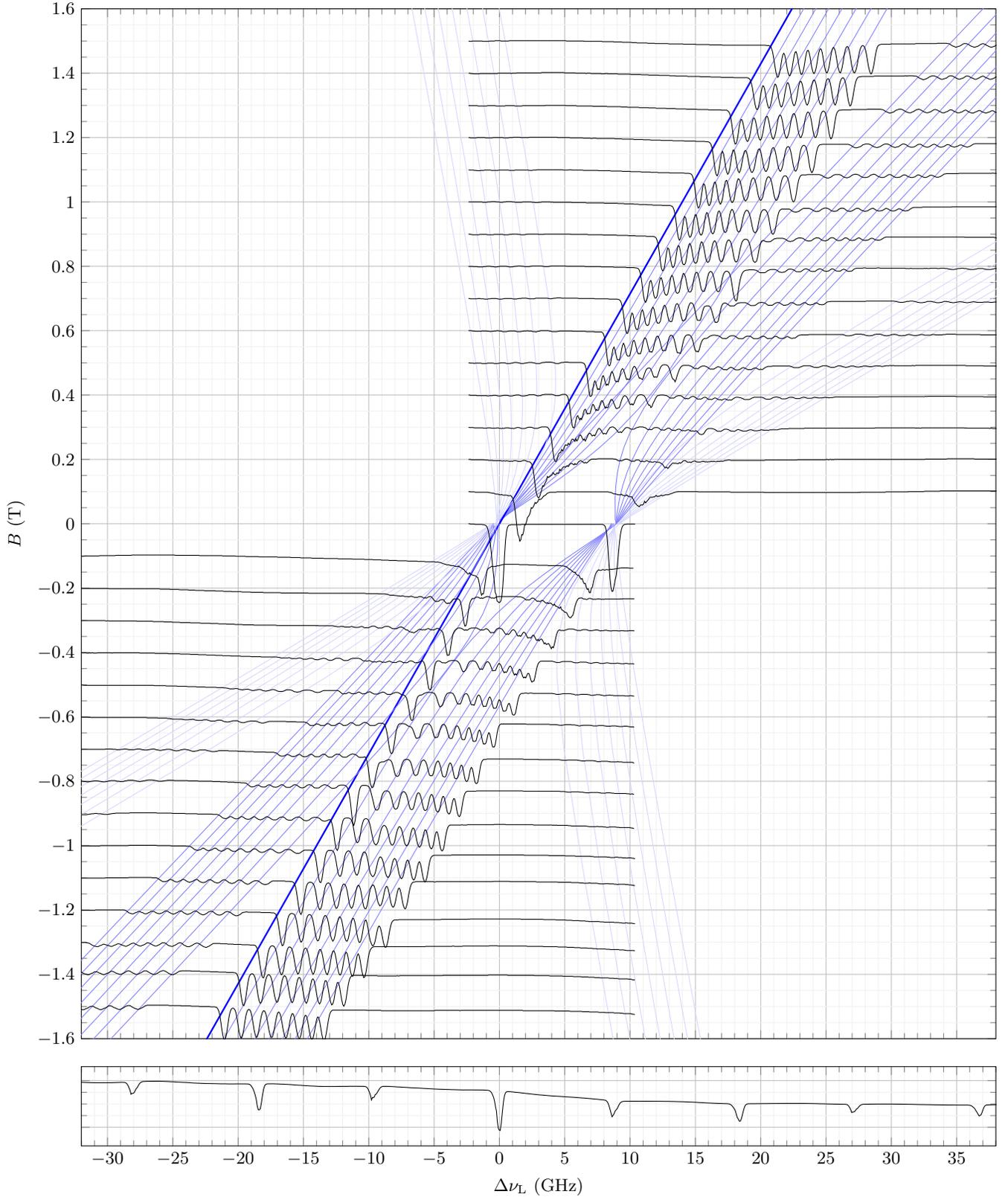
\begin{figure*}[p]
\centering
\begin{tikzpicture}
\begin{axis}[
			height              = 20 cm,
			width               = \textwidth,
			ylabel              = $B$ (T),
			xmin                = -32,
			xmax                = 38,
			xtick               = {-30,-25,-20,-15,-10,-5,0,5,10,15,20,25,30,35},
			xticklabels         = {$\textcolor{white}{-30}$,$\textcolor{white}{-25}$,$\textcolor{white}{-20}$,$\textcolor{white}{-15}$,$\textcolor{white}{-10}$,$\textcolor{white}{-5}$,$\textcolor{white}{0}$,$\textcolor{white}{5}$,$\textcolor{white}{10}$,$\textcolor{white}{15}$,$\textcolor{white}{20}$,$\textcolor{white}{25}$,$\textcolor{white}{30}$,$\textcolor{white}{35}$},
			ymin                = -1.6,
			ymax                = 1.6,
			grid                = both,
			grid style          = {line width=.4pt, draw=gray!10},
            major grid style    = {line width=.4pt, draw=gray!50},
			minor x tick num    = 4,
			minor y tick num    = 3,
			clip = true]
\ifNotDraft
\addplot [color=my_lighter_blue] table[y=B,x=1] {Intermediate_Field_Spectra/Lines_2_T.txt};
\addplot [color=my_lighter_blue] table[y=B,x=2] {Intermediate_Field_Spectra/Lines_2_T.txt};
\addplot [color=my_lighter_blue] table[y=B,x=3] {Intermediate_Field_Spectra/Lines_2_T.txt};
\addplot [color=my_lighter_blue] table[y=B,x=4] {Intermediate_Field_Spectra/Lines_2_T.txt};
\addplot [color=my_lighter_blue] table[y=B,x=5] {Intermediate_Field_Spectra/Lines_2_T.txt};
\addplot [color=my_lighter_blue] table[y=B,x=6] {Intermediate_Field_Spectra/Lines_2_T.txt};
\addplot [color=my_lighter_blue] table[y=B,x=7] {Intermediate_Field_Spectra/Lines_2_T.txt};

\addplot [color=my_light_blue] table[y=B,x=8] {Intermediate_Field_Spectra/Lines_2_T.txt};
\addplot [color=my_light_blue] table[y=B,x=9] {Intermediate_Field_Spectra/Lines_2_T.txt};
\addplot [color=my_light_blue] table[y=B,x=10] {Intermediate_Field_Spectra/Lines_2_T.txt};
\addplot [color=my_light_blue] table[y=B,x=11] {Intermediate_Field_Spectra/Lines_2_T.txt};
\addplot [color=my_light_blue] table[y=B,x=12] {Intermediate_Field_Spectra/Lines_2_T.txt};
\addplot [color=my_light_blue] table[y=B,x=13] {Intermediate_Field_Spectra/Lines_2_T.txt};
\addplot [color=my_light_blue] table[y=B,x=14] {Intermediate_Field_Spectra/Lines_2_T.txt};
\addplot [color=my_light_blue] table[y=B,x=15] {Intermediate_Field_Spectra/Lines_2_T.txt};

\addplot [color=my_light_blue] table[y=B,x=16] {Intermediate_Field_Spectra/Lines_2_T.txt};
\addplot [color=my_light_blue] table[y=B,x=17] {Intermediate_Field_Spectra/Lines_2_T.txt};
\addplot [color=my_light_blue] table[y=B,x=18] {Intermediate_Field_Spectra/Lines_2_T.txt};
\addplot [color=my_light_blue] table[y=B,x=19] {Intermediate_Field_Spectra/Lines_2_T.txt};
\addplot [color=my_light_blue] table[y=B,x=20] {Intermediate_Field_Spectra/Lines_2_T.txt};
\addplot [color=my_light_blue] table[y=B,x=21] {Intermediate_Field_Spectra/Lines_2_T.txt};
\addplot [color=my_light_blue] table[y=B,x=22] {Intermediate_Field_Spectra/Lines_2_T.txt};
\addplot [color=blue, line width = 0.3 mm]      table[y=B,x=23] {Intermediate_Field_Spectra/Lines_2_T.txt};

\addplot [color=my_lighter_blue] table[y=B,x=24] {Intermediate_Field_Spectra/Lines_2_T.txt};
\addplot [color=my_lighter_blue] table[y=B,x=25] {Intermediate_Field_Spectra/Lines_2_T.txt};
\addplot [color=my_lighter_blue] table[y=B,x=26] {Intermediate_Field_Spectra/Lines_2_T.txt};
\addplot [color=my_lighter_blue] table[y=B,x=27] {Intermediate_Field_Spectra/Lines_2_T.txt};
\addplot [color=my_lighter_blue] table[y=B,x=28] {Intermediate_Field_Spectra/Lines_2_T.txt};
\addplot [color=my_lighter_blue] table[y=B,x=29] {Intermediate_Field_Spectra/Lines_2_T.txt};
\addplot [color=my_lighter_blue] table[y=B,x=30] {Intermediate_Field_Spectra/Lines_2_T.txt};

\addplot [color=my_lighter_blue] table[y=-B,x=31] {Intermediate_Field_Spectra/Lines_2_T.txt};
\addplot [color=my_lighter_blue] table[y=-B,x=32] {Intermediate_Field_Spectra/Lines_2_T.txt};
\addplot [color=my_lighter_blue] table[y=-B,x=33] {Intermediate_Field_Spectra/Lines_2_T.txt};
\addplot [color=my_lighter_blue] table[y=-B,x=34] {Intermediate_Field_Spectra/Lines_2_T.txt};
\addplot [color=my_lighter_blue] table[y=-B,x=35] {Intermediate_Field_Spectra/Lines_2_T.txt};
\addplot [color=my_lighter_blue] table[y=-B,x=36] {Intermediate_Field_Spectra/Lines_2_T.txt};
\addplot [color=my_lighter_blue] table[y=-B,x=37] {Intermediate_Field_Spectra/Lines_2_T.txt};

\addplot [color=my_light_blue] table[y=-B,x=38] {Intermediate_Field_Spectra/Lines_2_T.txt};
\addplot [color=my_light_blue] table[y=-B,x=39] {Intermediate_Field_Spectra/Lines_2_T.txt};
\addplot [color=my_light_blue] table[y=-B,x=40] {Intermediate_Field_Spectra/Lines_2_T.txt};
\addplot [color=my_light_blue] table[y=-B,x=41] {Intermediate_Field_Spectra/Lines_2_T.txt};
\addplot [color=my_light_blue] table[y=-B,x=42] {Intermediate_Field_Spectra/Lines_2_T.txt};
\addplot [color=my_light_blue] table[y=-B,x=43] {Intermediate_Field_Spectra/Lines_2_T.txt};
\addplot [color=my_light_blue] table[y=-B,x=44] {Intermediate_Field_Spectra/Lines_2_T.txt};
\addplot [color=blue, line width = 0.3 mm]      table[y=-B,x=45] {Intermediate_Field_Spectra/Lines_2_T.txt};

\addplot [color=my_light_blue] table[y=-B,x=46] {Intermediate_Field_Spectra/Lines_2_T.txt};
\addplot [color=my_light_blue] table[y=-B,x=47] {Intermediate_Field_Spectra/Lines_2_T.txt};
\addplot [color=my_light_blue] table[y=-B,x=48] {Intermediate_Field_Spectra/Lines_2_T.txt};
\addplot [color=my_light_blue] table[y=-B,x=49] {Intermediate_Field_Spectra/Lines_2_T.txt};
\addplot [color=my_light_blue] table[y=-B,x=50] {Intermediate_Field_Spectra/Lines_2_T.txt};
\addplot [color=my_light_blue] table[y=-B,x=51] {Intermediate_Field_Spectra/Lines_2_T.txt};
\addplot [color=my_light_blue] table[y=-B,x=52] {Intermediate_Field_Spectra/Lines_2_T.txt};
\addplot [color=my_light_blue] table[y=-B,x=53] {Intermediate_Field_Spectra/Lines_2_T.txt};

\addplot [color=my_lighter_blue] table[y=-B,x=54] {Intermediate_Field_Spectra/Lines_2_T.txt};
\addplot [color=my_lighter_blue] table[y=-B,x=55] {Intermediate_Field_Spectra/Lines_2_T.txt};
\addplot [color=my_lighter_blue] table[y=-B,x=56] {Intermediate_Field_Spectra/Lines_2_T.txt};
\addplot [color=my_lighter_blue] table[y=-B,x=57] {Intermediate_Field_Spectra/Lines_2_T.txt};
\addplot [color=my_lighter_blue] table[y=-B,x=58] {Intermediate_Field_Spectra/Lines_2_T.txt};
\addplot [color=my_lighter_blue] table[y=-B,x=59] {Intermediate_Field_Spectra/Lines_2_T.txt};
\addplot [color=my_lighter_blue] table[y=-B,x=60] {Intermediate_Field_Spectra/Lines_2_T.txt};

\addplot [color=black] table[x=Freq,y=Spectrum] {Intermediate_Field_Spectra/Spectrum_at_+0.0_T.txt};
\addplot [color=black] table[x=Freq,y=Spectrum] {Intermediate_Field_Spectra/Spectrum_at_+0.1_T.txt};
\addplot [color=black] table[x=Freq,y=Spectrum] {Intermediate_Field_Spectra/Spectrum_at_+0.2_T.txt};
\addplot [color=black] table[x=Freq,y=Spectrum] {Intermediate_Field_Spectra/Spectrum_at_+0.3_T.txt};
\addplot [color=black] table[x=Freq,y=Spectrum] {Intermediate_Field_Spectra/Spectrum_at_+0.4_T.txt};
\addplot [color=black] table[x=Freq,y=Spectrum] {Intermediate_Field_Spectra/Spectrum_at_+0.5_T.txt};
\addplot [color=black] table[x=Freq,y=Spectrum] {Intermediate_Field_Spectra/Spectrum_at_+0.6_T.txt};
\addplot [color=black] table[x=Freq,y=Spectrum] {Intermediate_Field_Spectra/Spectrum_at_+0.7_T.txt};
\addplot [color=black] table[x=Freq,y=Spectrum] {Intermediate_Field_Spectra/Spectrum_at_+0.8_T.txt};
\addplot [color=black] table[x=Freq,y=Spectrum] {Intermediate_Field_Spectra/Spectrum_at_+0.9_T.txt};
\addplot [color=black] table[x=Freq,y=Spectrum] {Intermediate_Field_Spectra/Spectrum_at_+1.0_T.txt};
\addplot [color=black] table[x=Freq,y=Spectrum] {Intermediate_Field_Spectra/Spectrum_at_+1.1_T.txt};
\addplot [color=black] table[x=Freq,y=Spectrum] {Intermediate_Field_Spectra/Spectrum_at_+1.2_T.txt};
\addplot [color=black] table[x=Freq,y=Spectrum] {Intermediate_Field_Spectra/Spectrum_at_+1.3_T.txt};
\addplot [color=black] table[x=Freq,y=Spectrum] {Intermediate_Field_Spectra/Spectrum_at_+1.4_T.txt};
\addplot [color=black] table[x=Freq,y=Spectrum] {Intermediate_Field_Spectra/Spectrum_at_+1.5_T.txt};
\addplot [color=black] table[x=Freq,y=Spectrum] {Intermediate_Field_Spectra/Spectrum_at_-0.1_T.txt};
\addplot [color=black] table[x=Freq,y=Spectrum] {Intermediate_Field_Spectra/Spectrum_at_-0.2_T.txt};
\addplot [color=black] table[x=Freq,y=Spectrum] {Intermediate_Field_Spectra/Spectrum_at_-0.3_T.txt};
\addplot [color=black] table[x=Freq,y=Spectrum] {Intermediate_Field_Spectra/Spectrum_at_-0.4_T.txt};
\addplot [color=black] table[x=Freq,y=Spectrum] {Intermediate_Field_Spectra/Spectrum_at_-0.5_T.txt};
\addplot [color=black] table[x=Freq,y=Spectrum] {Intermediate_Field_Spectra/Spectrum_at_-0.6_T.txt};
\addplot [color=black] table[x=Freq,y=Spectrum] {Intermediate_Field_Spectra/Spectrum_at_-0.7_T.txt};
\addplot [color=black] table[x=Freq,y=Spectrum] {Intermediate_Field_Spectra/Spectrum_at_-0.8_T.txt};
\addplot [color=black] table[x=Freq,y=Spectrum] {Intermediate_Field_Spectra/Spectrum_at_-0.9_T.txt};
\addplot [color=black] table[x=Freq,y=Spectrum] {Intermediate_Field_Spectra/Spectrum_at_-1.0_T.txt};
\addplot [color=black] table[x=Freq,y=Spectrum] {Intermediate_Field_Spectra/Spectrum_at_-1.1_T.txt};
\addplot [color=black] table[x=Freq,y=Spectrum] {Intermediate_Field_Spectra/Spectrum_at_-1.2_T.txt};
\addplot [color=black] table[x=Freq,y=Spectrum] {Intermediate_Field_Spectra/Spectrum_at_-1.3_T.txt};
\addplot [color=black] table[x=Freq,y=Spectrum] {Intermediate_Field_Spectra/Spectrum_at_-1.4_T.txt};
\addplot [color=black] table[x=Freq,y=Spectrum] {Intermediate_Field_Spectra/Spectrum_at_-1.5_T.txt};
\fi
\end{axis}
\end{tikzpicture}
\begin{tikzpicture}
\begin{axis}[
			height              = 3 cm,
			width               = \textwidth,
			ylabel              = $\textcolor{white}{B (T)}$,
			xlabel              = $\Delta\nu_\mathrm{L}$ (GHz),
			xmin                = -32,
			xmax                = 38,
			ymin                = -1.4,
			ymax                = 0.3,
			ytick               = {0,-1},
			yticklabels         = {$\textcolor{white}{0.0}$,$\textcolor{white}{-1.0}$},
			grid                = both,
			grid style          = {line width=.4pt, draw=gray!10},
            major grid style    = {line width=.4pt, draw=gray!50},
			minor x tick num    = 4,
			minor y tick num    = 3,
			clip                = true]
\ifNotDraft
\addplot [color=black] table[x=Freq,y=Spectrum] {Intermediate_Field_Spectra/Spectrum_reference.txt};
\fi
\end{axis}
\end{tikzpicture}
\captionsetup{width=10 cm}
\caption{Experimental verification of the line splitting, using the increasing field in front of the 7 T scanner. In the lower plot the reference probe, at 0 T, having sidebands of 18.386 GHz, is used to estimate the frequency axis, assuming a linear laser frequency scan $\Delta\nu_\mathrm{L}$. The calculated lines, are overlaid on top of the spectra. Notice that the axes are flipped, compared to \cref{fig:Lines_splitting_negative}. We see a good agreement with the calculated lines.}
\label{fig:Intermediate_field_spectra}
\end{figure*}

The transition from the Zeeman regime to the hyperfine Paschen-Back regime for  $\sigma_+$ lines, as shown in \cref{fig:Lines_splitting_negative} is experimentally verified by using the increasing magnetic field in front of the 7 T scanner. This is shown in \cref{fig:Intermediate_field_spectra}, with the theoretical lines overlaid. Notice that the axes are flipped, compared to \cref{fig:Lines_splitting_negative}. 

The laser scan is mapped to a frequency axis by using the reference probe in the magnetic shield, with sidebands of 18.386 GHz, as shown in the bottom of the plot. A linear scan is assumed. Notice that for these measurements the EOM shown in \cref{fig:Setup} is moved to the beam path going to the reference, instead of probe 1-4.

Since the laser scan is limited to about 20 GHz, each line is actually two scans stitched together (and the reference plot is made from four scans). As the field strengths are approximate, and the frequency axis is only approximately linear, this data should only be appreciated qualitatively.

\bibliography{Bibliography}

\makeatletter\onecolumngrid@push\makeatother 
\clearpage 
\makeatletter\onecolumngrid@pop\makeatother 

\end{document}